\documentclass[12pt,english]{article}
\pdfoutput=1
\usepackage{jheppub}
\usepackage{simplewick}
\usepackage{hyperref}
\usepackage{verbatim}
\usepackage{graphics, color}
\usepackage{graphicx}
\usepackage{amsmath}
\usepackage{amssymb}
\usepackage{amsthm}
\usepackage{amsfonts}
\usepackage[usenames,dvipsnames]{xcolor}
\usepackage[normalem]{ulem}
\usepackage{braket}
\usepackage{url}
\usepackage{subfig}

\begin{document}

\title{Operator Size for Holographic Field Theories}
\author{Alexandros Mousatov$^1$}
\affiliation{$^1$Stanford Institute for Theoretical Physics, Department of Physics, Stanford University, Stanford, CA 94305, USA}
\emailAdd{mousatov@stanford.edu}
\abstract{
We formulate a state-dependent definition of operator size that captures the effective size of an operator acting on a reference state. We apply our definition to the SYK model and holographic 2-dimensional CFTs, generalizing the Qi-Streicher formula to a large class of geometries which includes pure AdS$_3$ and BTZ black holes. In vacuum AdS$_3$, the operator size is proportional to the global Hamiltonian at leading order in $1/N$, mirroring the results of Lin-Maldacena-Zhao in AdS$_2$. For BTZ geometries, it is given by the sum of the Kruskal momenta. Higher $1/N$ corrections become relevant when backreaction gets large, and we expect a transition in the growth pattern that depends on the transverse profile of the excitation. We propose a bulk dual that captures this profile dependence and exhibits saturation at a size of order the black hole entropy. This bulk dual is an averaged eikonal phase over a class of scattering events, and it can be interpreted as the ``number of virtual gravitons" in the gravitational field created by an infaller.

}

\maketitle
\flushbottom
\addtocontents{toc}{\protect\enlargethispage{2\baselineskip}}

\newpage

\section{Introduction}

	Around fifty years ago, the renormalization group provided a new insight into the dynamics of quantum many-body systems, demonstrating that long-distance physics can be universal regardless of the microscopic physics that occurs at the lattice scale. In recent work on chaotic systems, a new notion of universality appears to emerge from the notion of operator growth. In holographic theories, this first manifested in the spreading of entanglement \cite{ButterflyEffect, EntanglementTsunami} and the growth pattern of OTOCs \cite{StringyEffects}, which have been shown to grow exponentially with a maximal Lyapunov exponent $\lambda_L = \frac{2\pi}{\beta}$\cite{ChaosBound}. In the limit where $\beta \ll t \ll \beta \log N$, the OTOC dynamics of large-$N$ theories are dominated by a hydrodynamic mode \cite{SoftModeSYK, HydroCloud, AdS3Reparametrizations, ChaosEFT,CFTQuantumChaos} whose bulk description is given by gravitational interactions. Any other interaction is subleading at this stage, and all one needs to know to compute the OTOCs $\langle V(t)WV(t)W\rangle_\beta$ are the couplings to this hydrodynamic mode. 

The OTOCs in a large-$N$ theory are thought to be a proxy for the ``size" of an operator, a connection that has been made precise for SYK-like systems in \cite{OperatorGrowth, Streicher-Qi}. In the regime $\beta \ll t \ll \beta \log N$, operators grow exponentially with a rate that's proportional to their current size
\begin{equation}
\frac{ds}{dt} \simeq \lambda_L s
\end{equation}
The universality for the growth of OTOCs can thus be interpreted as a universality for the growth of complex operators. For theories with many degrees of freedom, the dynamics of sufficiently complex operators are self-averaging, and thus they can be described semi-classically. In the bulk of a holographic theory, this semi-classical description is given by gravity. For example, in the limit where SYK is well-described by the Schwarzian theory \cite{MaldacenaStanfordYang}, the size of a Majorana fermion $\psi_R(-t)$ is proportional to the strength of the shockwave it creates in the bulk \cite{Streicher-Qi}. The strength of the shockwave dictates the gravitational dynamics that control geodesic lengths in the bulk, and on the boundary the size of $\psi_R(-t)$ controls the OTOCs $\langle \psi_R(-t) \psi_L \psi_R \psi_R(-t)\rangle_\beta$.

In \cite{ThingsFall, FallingCharged}, a simple formula was proposed that related the radial momentum of a bulk particle with the size of its dual operator. The strength of a shockwave (as measured by the time-delay it causes) is directly proportional to the radial momentum, and thus the operator size distribution of precursors in SYK was found to match precisely the radial momentum wavefunction of a bulk particle in AdS$_2$ \cite{Streicher-Qi}, providing strong evidence for the correspondence. Our goal in this paper is to investigate the holographic dual of operator size for more general geometries, both in AdS$_2$ and in higher dimensional spacetimes.

While the notion of operator size has been made precise in the context of the SYK model \cite{OperatorGrowth, Streicher-Qi} and many body systems with $q$-level sites \cite{LucasThermal}, these definitions apply only to thermal ensembles of finite-dimensional theories\footnote{The Qi-Streicher formula \cite{Streicher-Qi} technically applies to density matrices $\rho$, but as we will see the definition becomes ill-behaved for pure states.}. A state-dependent definition of size that applies to pure states and higher-dimensional field theories (which have an infinite-dimensional Hilbert space even at finite spatial extent) is needed to investigate the size-momentum correspondence in larger generality. 

In this paper, we formulate such a definition, and then we use it to investigate operator size in a large class of holographic systems. While in SYK it is possible to find a ``microscopic" definition of operator size, we haven't been able to do so for higher dimensional field theories. Instead, we have followed a phenomenological method. Starting from the base assumption that OTOCs must measure the operator size in some sense (which we will make more precise), we have found that operator size can be uniquely determined in highly symmetric states (e.g. the AdS vacuum and its quotients). For convenience, we summarize our results and the organization of the paper in the next subsection.

\subsection{Summary of Results}

In this paper, we define the state-dependent size of an operator $\mathcal{O}$ by counting the average number of ``fundamental operators" $\mathcal{O}_i$ that must act on a reference state $|\Psi\rangle$ to yield $\mathcal{O}|\Psi\rangle$. This definition is highly reminiscent of complexity, except that we allow both sums and products of fundamental operators. In Section \ref{Section2} we will show that this definition can be made rigorous, and it yields an expression
\begin{equation}\label{IntroSize}
S_{|\Psi\rangle}(\mathcal{O}) = \frac{\langle \Psi|\mathcal{O}^\dagger \hat{S}_{|\Psi\rangle} \mathcal{O}|\Psi\rangle}{\langle \Psi| \mathcal{O}^\dagger \mathcal{O} |\Psi\rangle}
\end{equation}
for some positive semi-definite, Hermitian operator $\hat{S}_{|\Psi\rangle}$. Our construction of $\hat{S}_{|\Psi\rangle}$ can be performed numerically in small systems, but in general this is more of an existence proof. The non-trivial aspect of the above equation is that the operator size can be written as a positive observable with a gapped spectrum (an energy of sorts), which is suggestive of its bulk interpretation.

In Section \ref{SYKModel}, we will use Equation \ref{IntroSize} as an ansatz to explicitly determine $\hat{S}_{|\Psi\rangle}$ in a large class of SYK states, including the thermofield double $|\beta\rangle$, the time-shifted thermofield doubles $|\beta(t_L, t_R)\rangle$, and the duals of single-sided black holes \cite{MaldacenaKourkoulou}. As a sanity check, we find that our result for $\hat{S}_{|\beta\rangle}$ is equivalent to the Qi-Streicher formula \cite{Streicher-Qi} in the large-$N$ limit\footnote{Both the Qi-Streicher formula and our own are technically only valid to leading order in $1/N$. As we will see in Section \ref{Section2}, they can give nonsensical results when applied to complex projection operators, but they are valid as long as we only care about relatively simple operators (i.e. with size $\ll N$).}. 

In Section \ref{FreeFieldSize} we will determine the operator $\hat{S}$ for free field theories, which will serve as a warmup for our attempt to formulate operator size in holographic field theories. This attempt will be the subject of Section \ref{CFTSection}, which forms the majority of this paper. Due to the difficulty of deriving a ``microscopic" definition of operator size, we will follow a more phenomenological route. Starting from the assumption that the growing part of OTOCs measures some part of an operator's size, we will try to construct the operator size $\hat{S}_{|\Omega\rangle}$ from operators that appear in the part of the OPE that is responsible for OTOC growth (or from the bulk perspective, for shockwave calculations \cite{ShockwaveOPE, EntropyEoM}). At leading order in $1/N$, the symmetries of the vacuum state uniquely determine its operator size to be proportional to the CFT Hamiltonian
\begin{equation}\label{IntroHoloSize}
\hat{S}_{|\Omega\rangle} = l_{AdS} H_{CFT}
\end{equation}
While this result may be surprising as it predicts no relation between the operator size of a particle and its position in the bulk, its consequences are much more in line with conventional wisdom\footnote{While the above result is valid for CFTs in any dimension, we will work with 2-dimensional CFTs in this paper. The fact that we can write any static 3-dimensional geometry as a quotient of pure AdS makes 3-dimensions much more tractable.}. First, by using a kinematic space formula \cite{KinematicComplexity} and the first law of entanglement entropy \cite{RelativeEntropyBulk} we find that at linearized order the operator size is proportional to the complexity increase (as suggested by the Complexity = Volume conjecture)

\begin{equation}
\hat{S}_{|\Omega\rangle} \sim \frac{\delta V}{l_{AdS} G_N}
\end{equation}
This proportionality was anticipated in \cite{ComplexityLaws} to hold for operators that are still far from saturation. In SYK, the same proportinality stems from the fact that geodesic lengths control both OTOCs and the maximal volume slice. The second consequence of Equation \ref{IntroHoloSize} is that it can be interpreted as an operator size for the AdS-Rindler thermofield double \cite{RindlerAdSCFT, RindlerQuantumGravity}. By taking quotients of AdS-Rindler to construct BTZ black holes, we find a universal formula for all non-rotating BTZ geometries, which in the near-horizon region reads
\begin{equation}
\hat{S}_{|BTZ\rangle} \sim P_u + P_v
\end{equation}
The operator size is given by the (dimensionless) Kruskal momenta, and it demonstrates the expected growth $e^{2\pi t/\beta}$. It also gives a clear demonstration of the momentum-size duality in higher dimensions. 

The above equations are only valid at leading order in $1/N$, and we expect them to be modified when backreaction is taken into account. In Section \ref{SectionBackreaction} we will investigate when these corrections become important, and we argue that they must lead to a transition in the growth pattern from exponential to power-law growth (which is given by groth rate of the ``butterfly cone") once backreaction becomes important. Interestingly, this transition can only happen for localized excitations; infalling spherical shells will keep growing exponentially until their size saturates. Any candidate operator size must be able to distinguish between these two scenarios.

In Section \ref{BulkSection} we propose a bulk quantity which exhibits these two distinct growth patterns (see Figures \ref{fig:localized}, \ref{fig:spherical}) for localized excitations and spherical shells. Furthermore, it exhibits a saturated value of order the black hole entropy for both types of shockwaves. This quantity is the average eikonal phase over a class of near-horizon scattering events that are dressed to the boundary (so that they can be defined despite backreaction). We find that these scattering events automatically stay at sub-Planckian energies, and they can thus be defined for much longer than the scrambling time.

Finally, in Section \ref{ComplexitySection} we comment on the universal behavior of operator size, complexity and a certain notion of ``average subsystem entropy" at linearized order in $1/c$. We attempt to explain their observed proportionality in terms of a toy model for the boundary CFT.

\section{State-Dependent Operator Size}\label{Section2}

In this section, we will provide a state-dependent definition of operator size for general quantum systems. Our state-dependent operator size should in some sense measure the number of fundamental operators that are necessary to reproduce the effect of $\mathcal{O}$ on a reference state $|\psi\rangle$. To demonstrate the subtleties that such a definition could introduce, consider an $N$-spin system in the state
\begin{equation}\label{SpinState}
|\psi\rangle = \frac{|\uparrow ... \uparrow\rangle + |\downarrow ... \downarrow\rangle}{\sqrt{2}}
\end{equation}
and we take the fundamental operators to include the Pauli matrices $\sigma_x^i$ acting on each spin $1, 2, ..., n$. Consider the operators built from $n$ Pauli matrices
\begin{equation}
\mathcal{O}_n = \prod_{i=1}^n \sigma_x^i
\end{equation}
For small $n < N/2$, it is clear that the size of $\mathcal{O}_n$ should be $n$, but for $n > N/2$, we have the ``shortcut"
\begin{equation}\label{SpinExample}
\mathcal{O}_n |\psi\rangle = \prod_{i=n+1}^N \sigma_x^i|\psi\rangle
\end{equation}
We can thus reproduce the effect of $O_n$ by acting on $|\psi\rangle$ with a shorter operator, and we'd expect that the state-dependent size of $\mathcal{O}_n$ should be $S_{|\psi\rangle}(\mathcal{O}_n) = N-n$. In the extreme case when $n = N$, we have $\mathcal{O}_N|\psi\rangle = |\psi\rangle$ and $\mathcal{O}_N$ acts as the identity operator, so its size is zero.

The above example illustrates that for certain states $|\psi\rangle$, an operator $\mathcal{O}$ may find its size to be shorter than what would be naively expected. Fundamentally, the problem is that each state $|\psi\rangle$ belongs to the kernel of various operators $A_i$, and thus these operators can provide ``shortcuts" that allow us to reproduce the effect of large operators with shorter ones. In the above case for example, $|\psi\rangle$ belongs to the kernel of the operators
\begin{equation}
\mathcal{O}_n - \prod_{i=n+1}^N \sigma^i_x
\end{equation}
which allows the substitution of $\mathcal{O}_n$ for a shorter operator whenever $n > N/2$. Thus, one has to quotient out by such substitutions before calculating an operator's size. 

In Section \ref{Review}, we will review the state-independent (and thermal ensemble) definitions of operator size, and explain why they do not readily generalize to pure states. Then in Section \ref{Construction} we will demonstrate how to correct this issue in fermionic systems and obtain an expression like in Equation \ref{IntroSize}. In Appendix \ref{Bosons} we deal with subtleties involving the infinite dimensional Hilbert spaces of bosonic systems, but the upshot is that size can still be written in the same way. The main difference is that a state-independent operator size cannot be defined, and it's necessary to write a state-dependent construction from the start. The reader can safely skip the Appendix on a first read, as the subtleties won't be particularly relevant in the rest of the paper.

\subsection{A Review of Operator Size}\label{Review}

For convenience, let us assume that we work with an SYK-esque system whose Hilbert space $H$ with dimension $|H| = 2^{N/2}$\footnote{Operator size has been defined for more general qudit systems in \cite{LucasThermal}	. Here, we will work with qubit systems for the sake of convenience.}. The Hilbert space $H$ then admits a tensor factorization into qubits, and we can generate the algebra of operators on $H$ from a set of $N$ Majorana fermions $\psi_i$ satisfying the canonical anti-commutation relations
\begin{equation}
\lbrace \psi_i, \psi_j \rbrace = 2\delta_{ij}
\end{equation}
For an operator $\mathcal{O} = \sum c_{i_1 ... i_k} \psi_{i_1}...\psi_{i_k}$ expanded in terms of Majorana fermions, the state-independent operator size is given by \cite{OperatorGrowth}
\begin{equation}\label{AverageSize}
S_{\infty}(\mathcal{O}) = \frac{\sum k|c_{i_1 ... i_k}|^2}{\sum |c_{i_1 ... i_k}|^2}
\end{equation}
For an appropriately normalized operator with $\sum |c_{i_1 ... i_k}|^2 = 1$, operator size is then given by a sum of squared commutators
\begin{equation}\label{SizeFormula}
S_{\infty}(\mathcal{O}) = \frac{1}{4}\sum_{i=1}^N 2^{-\frac{N}{2}} \text{tr}(\lbrace \mathcal{O}, \psi_i \rbrace^\dagger \lbrace \mathcal{O}, \psi_i\rbrace)
\end{equation}
A convenient way to rewrite this formula is to work in the doubled Hilbert space and use the infinite-temperature thermofield double $|I\rangle$, which can be defined as the unique state satisfying

\begin{equation}
(\psi_L^j + i \psi_R^j)|I\rangle = 0
\end{equation}
for all $j = 1, 2, ..., N$. This allows for a convenient rewriting of the operator size as
\begin{equation}\label{DoubleSidedFormula}
S_{\infty}(\mathcal{O}) = \frac{\langle I|\mathcal{O}_R^\dagger \hat{S}_\infty \mathcal{O}_R |I\rangle}{\langle I|\mathcal{O}_R^\dagger \mathcal{O}_R|I\rangle}
\end{equation}
where
\begin{equation}
\hat{S}_\infty = \sum_j \frac{i}{2}\psi^j_L \psi^j_R + \frac{N}{2}
\end{equation}
An equivalent way to rewrite this, which we will later find to be useful, is
\begin{equation}\label{FermionFormula}
\hat{S} = \sum_j \frac{1}{4}(\psi^j_L + i\psi^j_R)^\dagger (\psi^j_L + i \psi^j_R)
\end{equation}
This expression manifestly shows that $\hat{S}$ is a positive definite operator (as a sum of operators of the form $A_i^\dagger A_i$) and furthermore $\hat{S}|I\rangle = 0$. In fact, $|I\rangle$ is the unique state annihilated by $\hat{S}$. The property that $\hat{S}$ counts operator size ultimately follows from $\hat{S}|I\rangle = 0$ plus the commutation relations
\begin{equation}
[ \hat{S}, \psi^j_R ] = i\psi^j_L
\end{equation}
The definition proposed in \cite{Streicher-Qi} for the thermal size of an operator was
\begin{equation}\label{ThermalSize}
S_{\beta}(\mathcal{O}) = \frac{1}{\delta_\beta}(S_{\infty}(\mathcal{O}\rho_\beta^{1/2}) - S_{\infty}(\rho_\beta^{1/2}))
\end{equation}
where $\rho_\beta = e^{-\beta H}$ is the thermal density matrix and $\delta_\beta =  G(\beta/2)$ is a ``size renormalization factor" that ensures the thermal size of one fermion is 1. One can equivalently write
\begin{equation}\label{ThermalSize2}
S_{\beta}(\mathcal{O}) = \frac{1}{\delta_\beta}\frac{\langle \beta | \mathcal{O}_R^\dagger (\hat{S} - \langle \hat{S}\rangle_\beta)\mathcal{O}_R|\beta\rangle}{\langle \beta |\mathcal{O}_R^\dagger \mathcal{O}_R |\beta\rangle}
\end{equation}
The above definition of operator size works splendidly for measuring the effective size of a fermion $\psi(t)$ in SYK, and in fact it predicts precise agreement with the average momentum of its bulk wavefunction. Furthermore, as suggested in \cite{Streicher-Qi}, it appears to generalize naturally to more general density matrices than $\rho_\beta$. However, it has some caveats that hinder generalizations, especially when we wish to consider pure states.

First, let us note that $S_\beta(\mathcal{O})$ isn't positive definite. While it was shown by explicit computation that it is positive for SYK fermions, there is no reason to expect that \ref{ThermalSize} will be positive for arbitrary $\mathcal{O}$. As an example, since $\rho_\beta$ is invertible, we can take $\mathcal{O} = \rho_\beta^{-1/2}$ and then $S_\beta(\mathcal{O}) = -\delta_\beta^{-1} S_\infty(\rho_\beta^{1/2}) < 0$. Of course, that's an awfully complex operator, and one would expect that ``simple" operators (i.e. those that are made from an $O(1)$ number of Majorana fermions) will have positive thermal size. So while the thermal size appears to correctly capture the operator size of simple operators, it falls short of being a comprehensive definition.

For practical purposes, the above isn't much of a problem. A more severe caveat is that if we wish to generalize \ref{ThermalSize} to a pure state $|\Psi\rangle$, we need to replace $\rho_{\beta}$ with a projection operator $P_\Psi$. However, it is easy to derive from \ref{DoubleSidedFormula} that for any projection operator $P_\Psi = |\Psi\rangle\langle \Psi|$ and operator $\mathcal{O}$ with $\mathcal{O}|\Psi\rangle \neq 0$ we have
\begin{equation}
S_{\infty}(\mathcal{O}P_\Psi) = \frac{N}{2}
\end{equation}
Thus, the candidate state-dependent size $S_{P_{\Psi}}$ obtained from \ref{ThermalSize} by replacing $\rho_\beta \rightarrow P_\Psi$ is trivial,
\begin{equation}
S_{P_\Psi}(\mathcal{O}) = 0
\end{equation}
for all $\mathcal{O}|\Psi\rangle \neq 0$, and it is $-N/2$ when $\mathcal{O}|\Psi\rangle = 0$. This isn't a particularly helpful definition, which we would at the very least wish to be non-trivial and capable of capturing operator growth and ``shortcuts"  of the kind we demonstrated for the spin-state \ref{SpinState}.

\subsection{Operator Size for Fermionic Systems}\label{Construction}

As we discussed earlier, intuitively we want the state-dependent size $S_{|\Psi\rangle}(\mathcal{O})$ of an operator $\mathcal{O}$ to be the smallest number of fundamental operators that we must use to replicate the effect of $\mathcal{O}$ on $|\Psi\rangle$. The first thing to note is that this means all operators $\mathcal{O},\mathcal{O}'$ with
\begin{equation}
\mathcal{O}|\Psi\rangle = \mathcal{O}'|\Psi\rangle
\end{equation} 
will have the same size. If we call the equivalence class of such operator $C_\Psi(\mathcal{O})$, then up to a normalization, we expect the size $S_{|\Psi\rangle}(\mathcal{O})$ to be the minimum ``naive size" among all $\mathcal{O}'$ in $C_{\Psi}(\mathcal{O})$. Let's start by determining what the ``naive size" should be. We write out the ``wavefunction" of $\mathcal{O}$ as
\begin{equation}\label{Expansion}
\mathcal{O}= \sum_I c_I \Gamma_I 
\end{equation}
where $\Gamma_I$ is a collection of monomials in the Majorana fermions $\psi^j$. The appropriate normalization factor for this wavefunction is the 2-point function $\langle \Psi |\mathcal{O}^\dagger \mathcal{O}|\Psi\rangle$, which is the same if we replace $\mathcal{O}$ with any $\mathcal{O}' \in C_\Psi(\mathcal{O})$. We will thus take the monomials $\Gamma_I$ to be normalized as
\begin{equation}
\langle \Psi|\Gamma_I^\dagger \Gamma_I |\Psi\rangle = 1
\end{equation}
in Equation \ref{Expansion}. If a monomial $\Gamma_I$ annihilates $|\Psi\rangle$ and the above normalization is impossible, then we will use an arbitrary normalization, e.g. $\text{Tr}(\Gamma_I^\dagger \Gamma_I) = 1$. We expect these monomials to drop out since they don't change the effect of $\mathcal{O}$ on $|\Psi\rangle$. We then write the ``un-normalized naive size" as
\begin{equation}\label{UnnormalizedSize}
\tilde{S}_{|\Psi\rangle}(\mathcal{O}) = \sum_I |c_I|^2 k_I
\end{equation}
Here, $k_I$ is the degree of $\Gamma_I$ if $\Gamma_I|\Psi\rangle \neq 0$, and it is $0$ otherwise. 

Besides lacking a normalization, the above expression doesn't account for any ``shortcuts" that $\mathcal{O}$ may undergo that allow it to be written in terms of a smaller operator. To account for this effect, we take an infimum over the class $C_\Psi(\mathcal{O})$, and thus we write the operator size as
\begin{equation}\label{InfimumSize}
S_{|\Psi\rangle}(\mathcal{O}) = \frac{\inf_{\mathcal{O}' \in C_{\Psi}(\mathcal{O})} \tilde{S}_{|\Psi\rangle}(\mathcal{O}')}{\langle \Psi |\mathcal{O}^\dagger \mathcal{O} |\Psi\rangle}  
\end{equation}
The above formula is manifestly positive semi-definite (with equality iff $\lambda I \in C_{\Psi}(\mathcal{O})$), but it is also very unwieldy. Furthermore, it is far from clear how it could be related to a bulk observable such as the radial momentum. We thus seek to write it in the form

\begin{equation}\label{SizeDesiredForm}
S_{|\Psi\rangle}(\mathcal{O}) = \frac{\langle \Psi|\mathcal{O}^\dagger \hat{S}_{|\Psi\rangle} \mathcal{O}|\Psi\rangle}{\langle \Psi|\mathcal{O}^\dagger \mathcal{O}|\Psi\rangle}
\end{equation}
for a Hermitian and positive semi-definite operator $\hat{S}_{|\Psi\rangle}$. The advantage of such an expression is that it is manifestly the expectation value of an observable (i.e. a Hermitian operator). Furthermore, the positivity condition is suggestive that $\hat{S}_{|\Psi\rangle}$ will be some sort of gapped energy operator which has $|\Psi\rangle$ as its ground state. The gap arises from the fact that non-trivial operators should have a minimum size of 1 unless they overlap with the identity. In a holographic theory, if $|\Psi\rangle$ has a gravity dual with vanishing matter stress-energy tensor (e.g. the AdS vacuum, or a black hole in equilibrium), one could use Average Null Energy operators integrated along bulk geodesics would form a good ansatz for $\hat{S}_{|\Psi\rangle}$ (though that's far from an exhaustive list of the ``building blocks" we could use). 

To obtain Equation \ref{SizeDesiredForm} from Equation \ref{InfimumSize}, first note that the equivalence class $C_\Psi(\mathcal{O})$ is given by a hyperplane in the space of operators $L(H)$ acting on the Hilbert space $H$. This hyperplane is spanned by the annihilators of $|\Psi\rangle$ (which we will collectively call $A_\Psi$) and it passes through $\mathcal{O}$. The second thing to note is that $\tilde{S}_{|\Psi\rangle}$ defines an inner-product in the space of operators given by
\begin{equation}
\langle \mathcal{O}_1, \mathcal{O}_2 \rangle_{\tilde{S}} = \sum_I c_{I,1}^* c_{I,2} k_I
\end{equation}
where $c_{I,1}, c_{I,2}$ are the coefficients of $\mathcal{O}_1, \mathcal{O}_2$ in the expansion \ref{Expansion}.

The problem of minimizing $\tilde{S}_{|\Psi\rangle}(\mathcal{O}')$ in the class $C_\Psi(\mathcal{O})$ is thus equivalent to finding the point of minimum radius (in the $\tilde{S}_{|\Psi\rangle}$-product) on a hyperplane. This can be solved by projecting $\mathcal{O}$ onto the $\tilde{S}_{|\Psi\rangle}$-orthogonal complement of the annihilators $A_\Psi$.

More explicitly, we can follow a Gram-Schmidt process to write an $\tilde{S}_{|\Psi\rangle}$-orthogonal basis for $A_\Psi$ and then extend it to the entire space. We can then write
\begin{equation}
\mathcal{O} = \sum a_i A_i + b_j B_j
\end{equation}
where $A_i, B_j$ form an $\tilde{S}_{|\Psi\rangle}$-orthonormal basis and $A_i$ annihilate $|\Psi\rangle$. By varying $\mathcal{O}'$ in $C_\Psi(\mathcal{O})$, we can freely tune the coefficients $a_i$ but we can't change the $b_j$ coefficients. Thus, we minimize $\langle \mathcal{O}', \mathcal{O}'\rangle_{\tilde{S}}$ by setting all $a_i = 0$ and thus
\begin{equation}
S_\Psi(\mathcal{O}) = \frac{\sum |b_j|^2 k_j}{\langle \Psi| \mathcal{O}^\dagger \mathcal{O} |\Psi\rangle}
\end{equation}
The numerator in the above expression defines a positive semidefinite bilinear form on the linear space of equivalence classes $C_\Psi(\mathcal{O})$ (which is the quotient $L(H)/A_\Psi$) and thus it yields a positive-semidefinite inner product on the same space which vanishes only on $C_\Psi(\lambda I)$. The equivalence classes $C_\Psi(\mathcal{O})$ can be identified with the Hilbert space $H$, since $C_\Psi(\mathcal{O})$ is uniquely determined by the state $\mathcal{O}|\Psi\rangle$, and thus we obtain an inner product on the states $\mathcal{O}|\Psi\rangle$. By a standard theorem in linear algebra, any such inner product can be written in terms of a positive semi-definite Hermitian operator and thus we have
\begin{equation}\label{InnerProductForm}
S_{|\Psi\rangle}(\mathcal{O}) = \frac{\langle \Psi| \mathcal{O}^\dagger \hat{S}_{|\Psi\rangle} \mathcal{O}|\Psi\rangle}{\langle \Psi| \mathcal{O}^\dagger \mathcal{O} |\Psi\rangle}
\end{equation}

An unfortunate problem is that we only obtained an existence proof, but not a particularly useful way for constructing $\hat{S}_{|\Psi\rangle}$. In the next sections, we will address some special cases where $\hat{S}_{|\Psi\rangle}$ can be guessed from general principles, but let us make a short comment on numerics. For relatively small values of $N$, a brute force procedure can be used to determine $\hat{S}_{|\Psi\rangle}$. Given a state $|\Psi\rangle$, the first step would be to determine the compute the normalizations of all monomials $\Gamma_I$. There are $2^N$ such monomials, which will take $O(2^N)$ elementary operations, and this allows the determining of the ``naive size" $\tilde{S}_{|\Psi\rangle}$. The second step is to find the subspace of annihilators. This involves solving the equation $A|\Psi\rangle = 0$ for the matrix $A$; this is a hugely degenerate system of linear equations which can be solved to fix a single column of $A$, and is expected to require at most $O(2^{3N/2})$ elementary operations. Then, we can perform a Gram-Schmidt procedure, which involves $O(2^{3N})$ operations, in order to find the $\tilde{S}_{|\Psi\rangle}$-orthogonal complement of $A_\Psi$ and thus directly determine $\hat{S}_{|\Psi\rangle}$. The bottleneck comes from the Gram-Schmidt procedure and its $O(2^{3N})$ operations, but it should be numerically tractable for $N \sim 10-15$. We can do somewhat better if we only care about the size of a specific operator $\psi(-t)$ rather than the full matrix $\hat{S}_{|\Psi\rangle}$. Then, the Gram-Schmidt process is superfluous, and then the bottleneck comes from solving for the annihilators of $|\Psi\rangle$, which takes $O(2^{3N/2})$ operations and should be tractable for $N \sim 20-30$.

\subsection{Time Evolution}

We will eventually want to work with non-equilibrium states, so that a time-dependent definition of operator size becomes necessary. Our definition of operator size depends explicitly on the state $|\Psi\rangle$, and implicitly on the choice of fundamental fields $\phi_i$. If we want to compute the size of an operator $\mathcal{O}$ at time $t$, should we evolve the fundamental fields to be $\phi_i(t)$, the state to be $|\Psi(t)\rangle$, or both?

The answer depends on which picture of quantum mechanics we use. In the Schrodinger picture, we evolve the state $|\Psi(t)\rangle = e^{-iHt}|\Psi\rangle$ but the fundamental fields remain the same. It is important then that when looking at the time-evolution of $S_{|\Psi(t)\rangle}(\mathcal{O})$ we don't time-evolve $\mathcal{O}$, but we time-evolve the state $\mathcal{O}|\Psi\rangle$ as 
\begin{equation}
e^{-iHt}\mathcal{O}|\Psi\rangle = \mathcal{O}(-t)|\Psi(t)\rangle
\end{equation}
The operator size can be thought of a measure of difference between the state $|\Psi\rangle$ and $|\Phi\rangle = \mathcal{O}|\Phi\rangle$, so this time-evolution isn't surprising in the Schrodinger picture. We can then compute the operator size as
\begin{equation}
S_{|\Psi(t)\rangle}(\mathcal{O}) = \frac{\langle \Psi(t)|\mathcal{O}^\dagger(-t)\hat{S}_{|\Psi(t)\rangle} \mathcal{O}(-t)|\Psi(t)\rangle}{\langle \Psi(t)|\mathcal{O}^\dagger(-t)\mathcal{O}(-t)|\Psi(t)\rangle}
\end{equation}
It is important to note that in this expression, $\hat{S}_{|\Psi(t)\rangle}$ is built out of time-independent (Schrodinger picture) fields. We can then rewrite this formula in Heisenberg picture by writing out $|\Psi(t)\rangle = e^{-iHt}|\Psi\rangle$ and thus
\begin{equation}
S_{|\Psi(t)\rangle}(\mathcal{O}) = \frac{\langle \Psi|\mathcal{O}^\dagger \hat{S}_{|\Psi(t)\rangle}(t) \mathcal{O}|\Psi\rangle}{\langle \Psi|\mathcal{O}^\dagger \mathcal{O}|\Psi\rangle}
\end{equation}
In the Heisenberg picture, the size operator $\hat{S}_{|\Psi(t)\rangle}(t)$ is built of time-evolved fields that live on time $t$, but $\mathcal{O}$ and $|\Psi\rangle$ remain as they were. This expression amounts to time-evolving the fundamental fields, while keeping everything else the same. Throughout the rest of this work, we will find it convenient to work in the Heisenberg picture.

\section{Operator Size in the SYK Model}\label{SYKModel}

In this section, we will use Equation \ref{InnerProductForm} as our starting point to define operator size for some interesting classes of SYK states. While Equation \ref{InnerProductForm} doesn't say much about the exact form of $\hat{S}_{|\Psi\rangle}$, what we'll do is try to guess an ansatz for $\hat{S}_{|\Psi\rangle}$ and then try to fix its coefficients by demanding that it correctly computes the size of simple monomials. As a toy example, let us try to compute $\hat{S}_{|I\rangle}$ for the infinite temperature thermofield double.

Since we want $\hat{S}_{|I\rangle}$ to annihilate $|I\rangle$ and to be positive definite, we can write it in the form\footnote{This ansatz was also used in \cite{BuildTFD} to find a gapped Hamiltonian whose ground state is the TFD. Formally, our size operator is very similar, it is a ``Hamiltonian" which should have an $O(1)$ gap in its spectrum (the size of the smallest operator that's orthogonal to the identity). In the next section when we work out operator size for free fields, we find that our expressions are very similar to the TFD Hamiltonians of \cite{BuildTFD}. }
\begin{equation}
\hat{S}_{|I\rangle} = \sum_i A^\dagger_i A_i
\end{equation}
where the operators $A_i$ annihilate $|I\rangle$. Since $\psi_L^j + i \psi_R^j$ annihilate $|I\rangle$, this gives a natural candidate
\begin{equation}
\hat{S}_{|I\rangle} = \frac{1}{4}\sum_j (\psi^j_L - i\psi^j_R) (\psi^j_L + i \psi^j_R)
\end{equation}
As we saw in Section \ref{Review}, this was the operator that was used to compute the ``state-independent" operator size, which here we interpret as being a state-dependent operator size that corresponds to the state $|I\rangle$. It is easy to check explicitly that it correctly computes the size of all monomials. This follows explicitly from the equal-time commutation relations, plus the fact that we can use $(\psi_L+i \psi_R)|I\rangle = 0$ to write any monomial as a product of exclusively right (or left) fermions.

\subsection{Finite Temperature TFD}\label{TFDSize}

Let us now try to work out the operator size for the thermofield double
\begin{equation}
|\beta\rangle = \frac{1}{\sqrt{Z_\beta}}\sum e^{-\beta E/2} |E\rangle_L |E\rangle_R
\end{equation}
The annihilators of the TFD are given by the KMS conditions
\begin{equation}
(\psi_L(i\tau) + i\psi_R(i\beta - i\tau))|\beta\rangle = 0
\end{equation}
and the modular Hamiltonian $H_R - H_L$. It is hard to directly build a good candidate for $\hat{S}_{|\beta\rangle}$ from these operators. Instead, what we will do is try to use a combinatorial argument to derive the form of $\hat{S}_{|\beta\rangle}$ in the large-$N$ limit. Let's start with a generic expansion 
\begin{equation}
\hat{S}_{|\beta\rangle} = \sum c^{i_1 ... i_k}_{j_1 ... j_k} \psi^{i_1}_{j_1} ... \psi^{i_k}_{j_k}
\end{equation}
where the $i's$ run over flavors and the $j$'s run over $L, R$. Since a Majorana fermion squares to 1, there are no repetitions of $\psi$'s in this expansion. At large-$N$ and small relative boosts (e.g. when we are computing correlation functions of monomials at the same time), correlation functions factorize into products of diagonal 2-point functions. Thus, at leading order in $N$ each term in the expansion of $\hat{S}_{|\beta\rangle}$ must have exactly two copies of each flavor index so that its non-vanishing matrix elements are diagonal in flavor space. Since $(\psi^i_j)^2 = 1$, this means that only the combination $\psi^i_L \psi^i_R$ can appear.

Furthermore, we need $\hat{S}_{|\beta\rangle}$ to have a vanishing expectation value\footnote{It is hard to directly talk about a statement of the form ``$\hat{S}_{|\beta\rangle}$ annihilates $|\beta\rangle$ to first order in $1/N$", so instead we want it to have vanishing matrix elements for all simple monomials. In particular, this requires that it has a vanishing expectation value.}, so we can choose the expansion to be built from products of the combinations $i\psi^i_L \psi^i_R - \langle i\psi^i_L \psi^i_R\rangle_\beta$, where we added the $i$ to make these building blocks Hermitian. Thus, we can write
\begin{equation}
\hat{S}_{|\beta\rangle} = \sum_{i_1, ..., i_k} c_k \prod_{j=1}^k (i\psi^{i_j}_L \psi^{i_j}_R - \langle i\psi^{i_j}_L \psi^{i_j}_R\rangle)
\end{equation}
Due to the $O(N)$ symmetry of the model at large-$N$, the coefficients depend only on $k$ and not the flavor indices. Now, let us try to compute the size of a monomial $\psi^1_R ... \psi^n_R$. The indices $i_1, ... , i_k$ have no repetitions, so simple combinatorics gives us a size
\begin{equation}
\sum_k c_k {n \choose k} \langle 2i \psi_R \psi_L\rangle^k
\end{equation}
We wish this to be equal to $n$ for all values of $n$, which uniquely fixes $c_1 = 1/\langle 2i\psi_R \psi_L\rangle = -1/\langle 2i\psi_L \psi_R\rangle$ and $c_k = 0$ otherwise\footnote{There is an implicit assumption that there are no ``shortcuts" available to simple operators (i.e. no annihilators that would allow us to have a small size than naively expected for a monomial).}. This reproduces the Qi-Streicher formula
\begin{equation}\label{ThermalSize3}
\hat{S}_{|\beta\rangle} = \sum_{j=1}^N \frac{\langle i \psi^j_L \psi^j_R \rangle_\beta -i \psi^j_L \psi^j_R}{2\langle i \psi_L \psi_R \rangle_\beta}
\end{equation}
Our derivation was based on three facts: (i) operator size can be written in terms of an operator $\hat{S}_{|\beta\rangle}$, (ii) simple correlation functions factorize at leading order in $1/N$, (iii) the TFD has an approximate $O(N)$ symmetry. Its failure to be an ``exact" operator size comes from the failure of the second assumption, which fails when we consider arbitrary polynomials of order $N$ Majorana fermions.

As long as we only care about the behavior of $O(1)$-size operators, the above formula should be right. As far as precursors go, we expect that it captures the exponentially growing part of operator growth before finite-N effects become important and saturation begins to occur. If we work in the $\beta\mathcal{J} \gg 1$ regime where the Schwarzian gives a good approximation to the dynamics and the theory has (near) maximal Lyapunov exponent, then we can interpret operator growth in terms of bulk shockwaves. A precursor $\psi_R^j(-t)$ creates a shockwave which causes $\psi_R, \psi_L$ to decorrelate. Each correlation function $\langle \psi_L \psi_R\rangle$ will schematically decrease as
\begin{equation}
\frac{\langle \psi_R^j(-t)\psi^i_L \psi^i_R\psi_R^j(-t)\rangle_\beta}{\langle \psi^i_L \psi^i_R\rangle_\beta} \simeq 1 - \frac{e^{\frac{2\pi t}{\beta}}}{N}
\end{equation}
for $t < t_{scr}$, so the size of the precursor will be
\begin{equation}
S_{|\beta\rangle}(\psi^j(-t)) \sim e^{\frac{2\pi t}{\beta}}
\end{equation}
If we try to measure the operator size at $t \gtrsim t_{scr}$, we will have
\begin{equation}
\hat{S}_{|\beta\rangle}(\psi^j_R(-t)) \simeq N/2
\end{equation}
This arises from the fact that a strong shockwave will completely decorrelate $\psi^i_L$ and $\psi^i_R$, thus setting $\langle \psi^j_R(-t) i\psi^i_L \psi^i_R \psi^j_R(-t)\rangle \simeq 0$ and thus we have a contribution of 1/2 per fermion, adding to a total of $N/2$. While the above formula does indeed predict a saturation of operator size at the maximally scrambled value, we can't be completely certain that the behavior of operator size is accurate at times $t \lesssim t_{scr}$. In these times, the deviation from the exponentially growing behavior is significant (so finite $N$ effects are important), but the size hasn't yet completely saturated at $N/2$, and we have no rigorous control over this regime.

\subsection{Time Shifted Thermofield Doubles}

We will now derive the operator size for the time-shifted thermofield double
\begin{equation}
|\beta(t_L, t_R)\rangle = e^{-iH_L t_L - iH_R t_R}|\beta\rangle
\end{equation}
Since $H_R - H_L$ annihilates $|\beta\rangle$, the above state only depends on the sum $t_L + t_R$, but we will keep the times written separately to indicate that this is the state that corresponds to the boundary times $(t_L, t_R)$. In our derivation of Equation \ref{ThermalSize3}, we made no specific reference to the times $t_L, t_R$, and thus we can repeat the above derivation to write 
\begin{equation}\label{ShiftedTFDSize}
\hat{S}_{|\beta(t_L, t_R)\rangle} = \sum_{j=1}^N \frac{\langle i \psi^j_L(t_L) \psi^j_R(t_R) \rangle_\beta -i \psi^j_L(t_L) \psi^j_R(t_R)}{2\langle i  \psi^j_L(t_L) \psi^j_R(t_R) \rangle_\beta}
\end{equation}
We have written the above operator in the ``Heisenberg picture", which means that the size of $\mathcal{O}$ is computed as
\begin{equation}
\hat{S}_{|\beta(t_L, t_R)\rangle}(\mathcal{O}) = \frac{\langle \beta|\mathcal{O}^\dagger \hat{S}_{|\beta(t_L, t_R)\rangle} \mathcal{O}|\beta\rangle}{\langle \beta |\mathcal{O}^\dagger \mathcal{O}|\beta\rangle}
\end{equation}
We made this choice for convenience, since expectation values in $|\beta\rangle$ are easier to study directly than those in $|\beta(t_L, t_R)\rangle$.

The operator size for the time-shifted TFD can be computed in the Schwarzian theory, which gives the gravitational contribution to the 4-point function \cite{MaldacenaStanfordYang}
\begin{align}\label{Schwarzian4pt}
\frac{\langle V_1 W_3 V_2 W_4 \rangle_{\text{grav}}}{\langle V_1 V_2 \rangle \langle W_3 W_4 \rangle} = \frac{\Delta^2}{2\pi N}\Bigg \lbrace \Big(-2 + \frac{u_{12}}{\tan \frac{u_{12}}{2}}\Big) \Big(-2 + \frac{u_{34}}{\tan \frac{u_{34}}{2}}\Big) + \\
\frac{2\pi \Big(\sin(\frac{u_1 - u_2 + u_3 - u_4}{2})- \sin(\frac{u_1 + u_2 - u_3 - u_4}{2})\Big)}{\sin \frac{u_{12}}{2} \sin \frac{u_{34}}{2}} + \frac{2\pi u_{23}}{\tan \frac{u_{12}}{2}\tan\frac{u_{34}}{2}}\Bigg \rbrace
\end{align}
The above 4-point function was computed in Euclidean signature and the ordering $u_4 < u_2 < u_3 < u_1$. To compute the size of $\psi(-t)$, we analytically continue to Lorentzian signature with $u_1 = -i t_R + \pi, u_2 = i t_R, u_3 = -i t + \tau, u_4 = -i t - \tau$. The operator size is given by $-N$ times the above 4-point function, and if we keep the exponentially growing pieces we get
\begin{equation}\label{TimeShiftedSize}
\hat{S}_{|\beta(t_L, t_R)\rangle}(\psi(-t)) \sim \frac{\cosh(\frac{2\pi}{\beta}t)}{\cosh(\frac{2\pi}{\beta}t_R)}
\end{equation}
At $t_R = 0$, we get the same exponential growth. However, we note that the prefactor to the growth is in general dependent on $t_R$. In particular, it decays exponentially as $e^{-2\pi t_R/\beta}$ for fixed $t$ and $t_R \gg \beta$. This exponential decay has an analogue in higher dimensions. Consider the geodesic length $d(t_L, t_R)$ connecting the left and right boundary (at equal spatial coordinate) in a BTZ geometry, and perturb the geometry with a shockwave. The length variation of the geodesic is then \cite{ComplexityShockwaves}
\begin{equation}
\delta d(t_R,t_R) \sim e^{\frac{2\pi (t-t_R)}{\beta} }
\end{equation}
This weakening effect of the shockwave translates to a smaller prefactor $e^{-2\pi t_R/\beta}$ on the OTOCs, and it suggests a slower operator growth. In particular, this suggests an $O(1)$ size when $\Delta t \sim 2 t_R$, or equivalently when $t \sim -t_R$. At infinite temperature, this is a manifestation of the equality
\begin{equation}\label{ReflectionEquality}
\psi_R(t)|I\rangle = -i\psi_L(-t)|I\rangle
\end{equation}
At finite temperature, the equality isn't quite as straightforward, instead it is given by the KMS conditions
\begin{equation}\label{KMS}
\psi_R(t + i\tau)|\beta\rangle = -i\psi_L(-t+i\beta-i\tau)|\beta\rangle
\end{equation}
If we were to split $\psi$ into a low-energy component $\psi^{IR}$ with energies $E \ll 1/\beta$, and a high-energy component $\psi^{IV}$ with energies $E \gtrsim 1/\beta$, then the KMS conditions for the IR component become
\begin{equation}\label{ThermalReflection}
\psi_R^{IR}(t)|\beta\rangle \simeq -i\psi_L^{IR}(-t)|\beta\rangle
\end{equation}
Now, recall that the usual limit for shockwave geometries is to simultaneously take $E \rightarrow 0$ as $e^{2\pi t/\beta} \rightarrow \infty$ while keeping their product fixed. Thus, shockwave computations implicitly project on the IR components of operators, and they obey the simpler equality \ref{ThermalReflection}\footnote{This property was referred to as the ``entanglement reflection principle" in \cite{ComplexityShockwaves}.}. This means that if we seek to represent a shockwave created by the operator $\psi_R(-t)$ on the slice $(t_L, t_R)$, we have a potential ``shortcut" which reduces the size of $\psi_R(-t)$ compared to what is naively expected. This becomes most severe when $t = -t_R$, and we can simply represent $\psi_R(-t_R)$ as $\psi_L(t_R)$.

This doesn't fully resolve the problems of Equation \ref{TimeShiftedSize}; the Equation yields an even smaller size when $t \ll t_R$. In the case where $t_R \gg t \gg \beta$, this yields an operator size that's much smaller than 1. We interpret this to mean that the gravitational contribution to operator size is small in this regime, and the dominant contribution comes from the identity exchange, so the size remains nearly constant. The growth that comes from the Schwarzian degrees of freedom cancels with the ``shortcuts" that are available due to the KMS conditions. The thermofield double isn't a generic state (from the 2-sided perspective), it is a very fine-tuned state when it comes to its correlation functions, so these kind of shortcuts aren't too surprising. 

Now, we may ask what happens as we push $t_R$ towards the scrambling time. If we trust our formulas so far, then we would notice that Equation \ref{TimeShiftedSize} has an explicit dependence on $t$ rather than $t-t_R$, even as the TFD becomes a ``typical" state. By typical we mean that its off-diagonal correlation functions vanish. If we treat $\psi_L, \psi_R$ as different fields, then at early times we have an off-diagonal correlation function $\langle \psi_L \psi_R \rangle = O(1)$, so the TFD is initially not typical. But at late times, all simple left-right correlation functions die off. As far as we can only probe the state using simple correlation functions at leading order in $1/N$, the TFD becomes 
\begin{equation}
|\beta(t_R, t_R)\rangle \rightarrow \rho_{\beta, L} \otimes \rho_{\beta, R}
\end{equation}
This is a coarse-grained statement, of course the state remains pure. However, the purification of the simple (coarse) degrees of freedom are complex (fine) degrees of freedom that cannot be easily probed (in contrast with $t_L = t_R = 0$, when simple degrees of freedom on the two sides purify each other).

So, even though our state becomes typical, Equation \ref{TimeShiftedSize} retains memory of the TFD's state at $t_L = t_R = 0$ (by virtue of having an explicit dependence on $t$). This suggests that we shouldn't trust \ref{TimeShiftedSize} at this regime, and indeed there is good reason not to. The problem is that our method captures an effective size operator that is only valid at leading order in $1/N$. But when the correlation functions $\langle i\psi_L(t_L) \psi_R(t_R)\rangle_\beta$ become of order $1/N$ (i.e. when the TFD becomes typical), Equation \ref{ShiftedTFDSize} no longer captures the leading order behavior. 

To create an operator size for the late-time BTZ, we will use the mirror operators of \cite{MirrorOperators}. While mirror operators are usually used in the context of black hole microstates, formally all that is required is that the algebra of simple operators doesn't annihilate the state we're working with. In this case, since simple combinations of the operators $\psi^j_L(t_R), \psi^j_R(t_R)$ cannot annihilate $|\beta\rangle$ (due to the vanishing of off-diagonal 2-point functions and large-$N$ factorization), we can define a set of mirror operators $\tilde{\psi}^j_L(t_R), \tilde{\psi}^j_R(t_R)$. To a good approximation, these mirror operators are given by $\psi^j_R(-t_R), \psi^j_L(-t_R)$. We can use these operators to create an operator size
\begin{equation}\label{LateTimeSize}
\hat{S}_{|\beta(t_R, t_R)\rangle} = \sum_{j=1}^N \frac{\langle i \tilde{\psi}^j_L(t_R) \psi^j_L(t_R) \rangle_\beta -i \tilde{\psi}^j_L(t_R) \psi^j_L(t_R)}{2\langle i  \tilde{\psi}^j_L(t_R) \psi^j_L(t_R) \rangle_\beta} + \frac{\langle i \tilde{\psi}^j_R(t_R) \psi^j_R(t_R) \rangle_\beta -i \tilde{\psi}^j_R(t_R) \psi^j_R(t_R)}{2\langle i  \tilde{\psi}^j_R(t_R) \psi^j_R(t_R) \rangle_\beta}
\end{equation}
This operator size assigns a size of $n$ to each monomial of $n$ Majorana fermions $\psi^j_L(t_L), \psi^j_R(t_R)$. Given that we expect no ``shortcuts" involving simple operators (even if we mix left and right-sided operators), this operator size seems that is correctly counts the size of all simple operators. Furthermore, the size of an operator $\psi(t)$ only depends on the difference $t - t_R$, so Equation \ref{LateTimeSize} passes this sanity check.

One may raise the objection that Equation \ref{LateTimeSize} cannot be correct since we have doubled the number of Majorana fermions. The maximum value of Equation \ref{LateTimeSize} is indeed double of that of \ref{ThermalSize3}. However, we should recall that this is simply an approximate expression for simple operators, and we cannot trust it for sizes of order $N$. Mirror operators appear to duplicate the number of Majorana fermions, but that's a mirage that arises when we restrict our attention to simple degrees of freedom. In the next section, we will see that a similar construction can yield an operator size for typical single-sided black holes in SYK. 

\subsection{Pure State Black Holes}\label{PureSYKSection}

We will now formulate operator size for single-sided black holes, in particular for the states introduced in \cite{MaldacenaKourkoulou}. The starting point for their construction is to consider a boundary state $|B_s\rangle$ defined by the relation
\begin{equation}
(\psi^{k}+is_k\psi^{k+N/2})|B_s\rangle = 0, \quad k = 1, ..., N/2
\end{equation}
The structure of $|B_s\rangle$ is similar to that of $|I\rangle$, with the pairs of fermions $(\psi^{k}, s_k \psi^{k+N/2})$ playing the same role as $(\psi^k_R, \psi^k_L)$ in $|I\rangle$. This suggests the operator size
\begin{equation}
\hat{S}_{|B_s\rangle} = \frac{1}{4}\sum_{j=1}^{N/2} (\psi^j- is_j\psi^{j+N/2}) (\psi^j + i s_j\psi^{j+N/2})
\end{equation}
Just as with $|I\rangle$, it is easy to use commutation relations to establish that the above operator correctly counts the size of monomials. As suggested by their similarity to the infinite temperature TFD, the states $|B_s\rangle$ have an extremely high energy; one can create approximately thermal states via Euclidean time evolution

\begin{equation}\label{PureSYKState}
|B_s(\beta)\rangle = e^{-\beta H/2}|B_s\rangle
\end{equation}
This state's diagonal 2-point functions are exactly thermal at large-$N$,
\begin{equation}
\langle \psi^j(t_1)\psi^j(t_2) \rangle_{B_s(\beta)} = G_\beta(t_1 - t_2)
\end{equation}
but the off-diagonal components are non-thermal and equal to
\begin{equation}\label{OffDiag}
\langle \psi^j(t_1)\psi^{j+N/2}(t_2) \rangle_{B_s(\beta)} = - is_j G_\beta(t_1)G_\beta(t_2) + O(1/N)
\end{equation}

\begin{figure}
\begin{center}
\includegraphics[height=6cm]{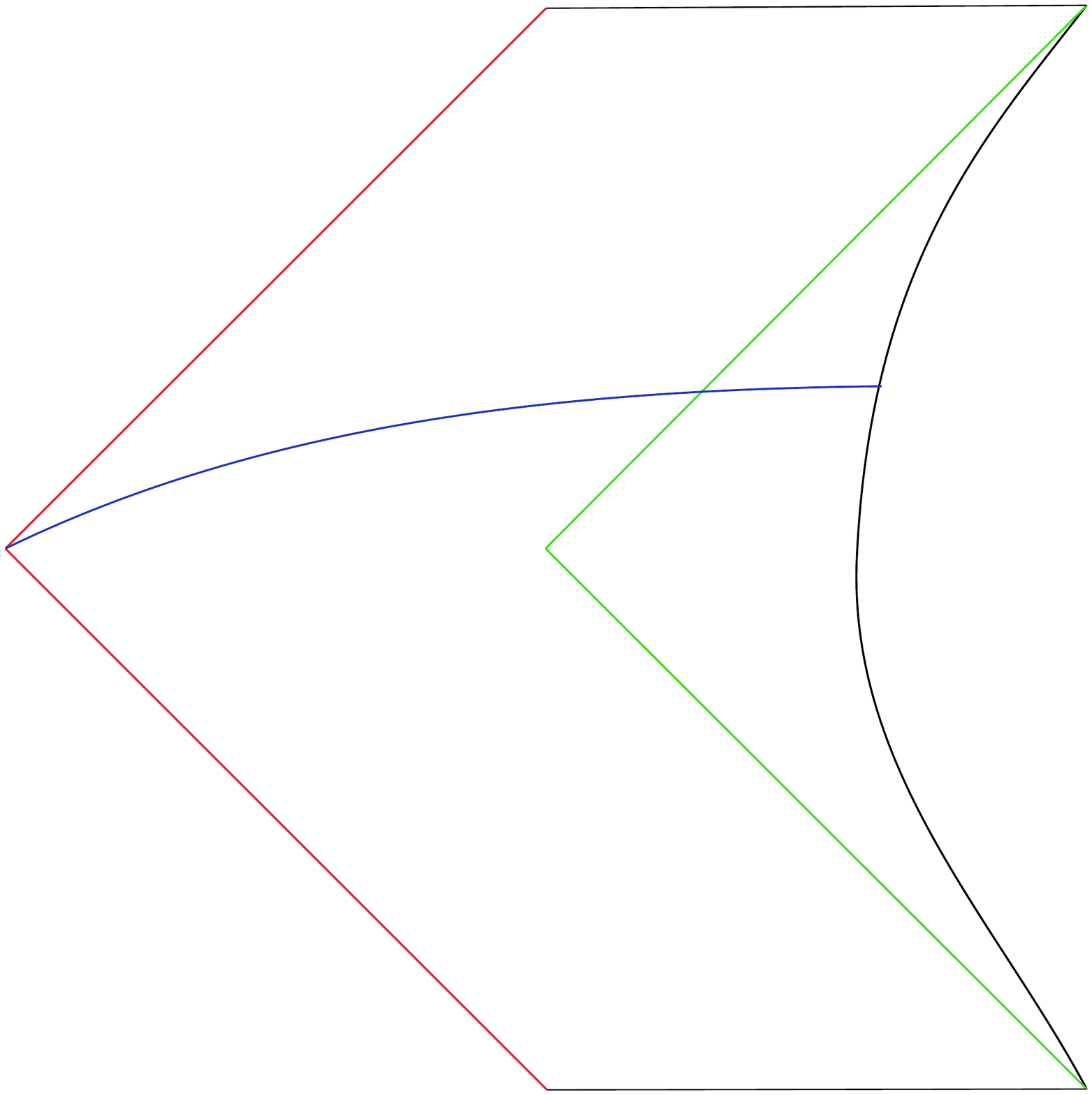}
\caption{The pure state \ref{PureSYKState} is dual to a single-sided black hole with a ETW brane (red) behind the horizon (green). The off-diagonal 2-point functions are given by powers of the length of the geodesic that connects a point on the boundary to the ``center" of the ETW brane.}\label{fig:SYKBH}
\end{center}
\end{figure}

The authors of \cite{MaldacenaKourkoulou} conjectured that the bulk dual of these states are single-sided black holes with an end-of-the-world brane (ETW brane) behind the horizon. The ETW brane has a boundary condition at $t = 0$ (see Figure \ref{fig:SYKBH}) that allows a $\psi^j$ bulk fermion to become a $\psi^{j+N/2}$ fermion. The off-diagonal 2-point function of $\psi^j(t_1) \psi^{j+N/2}(t_2)$ thus comes from a geodesic that goes from $t_1$ to the ETW brane, and then a geodesic that goes from the ETW brane to $t_2$.

The geometry is free of matter sources, and it satisfies the same factorization properties that we used to derive the operator size for the TFD. By using the same process as in Section \ref{TFDSize}, we can determine the operator $\hat{S}_{|B_s(\beta)\rangle}$ at leading order in $1/N$. This yields

\begin{equation}\label{BstateSize}
\hat{S}_{|\beta\rangle} = \sum_{j=1}^{N/2} \frac{\langle is_j \psi^j \psi^{j+N/2} \rangle_{B_s(\beta)} -i s_j \psi^j \psi^{j+N/2}}{2\langle is_j \psi^j \psi^{j+N/2}\rangle_{B_s(\beta)}}
\end{equation}
Using Equation \ref{OffDiag} we note that the expectation value $\langle is_j \psi^j \psi^{j+N/2}\rangle_{B_s(\beta)}$ is $s_j$ independent, and so we rewrite
\begin{equation}\label{BstateSize}
\hat{S}_{|B_s(\beta)\rangle} = \sum_{j=1}^{N/2} \frac{G_\beta(0)^2 -i s_j \psi^j \psi^{j+N/2}}{2G_\beta(0)^2}
\end{equation}
The exact same procedure can give the operator size for the time-shifted pure-state black hole $|B_s(\beta; t_R)\rangle$, we just need to replace $G_\beta(0) \rightarrow G_\beta(t_R)$. In the Schwarzian limit, we can relate the computation of a 4-point function $\langle\psi(-t)is_j \psi^j(t_R) \psi^{j+N/2}(t_R)\psi(-t)\rangle_{|B_s(\beta; t_R)\rangle}$ to computing the TFD correlation function

\begin{equation}
\frac{\langle \psi_R(-t) i\psi^j_L(0) \psi^j_R(t_R) \psi^{j+N/2}_L(0) \psi^{j+N/2}_R(t_R) \psi_R(-t)\rangle_\beta}{\langle i\psi^j_L(0) \psi^j_R(t_R)\rangle_\beta \langle i\psi^{j+N/2}_L(0) \psi^{j+N/2}_R(t_R)\rangle_\beta }
\end{equation}
To leading order in $1/N$, this a sum of Schwarzian 4-point functions given by Equation \ref{Schwarzian4pt}. This 4-point function is invariant under boosts, so we can evolve with $H_R - H_L$ to bring it to a symmetric configuration $\langle \psi_R(-t-t_R/2)\psi^j_L(t_R/2)\psi^j_R(t_R/2)\psi_R(-t-t_R/2)$. Then we obtain the result
\begin{equation}\label{PureStateSize}
S_{|B_s(\beta; t_R)\rangle}(\psi(-t)) \sim \frac{\cosh(\frac{2\pi}{\beta}(t-\frac{t_R}{2}))}{\cosh(\frac{2\pi}{\beta}\frac{t_R}{2})}
\end{equation}
The growth is (at leading order) identical with what we would obtain if we considered the asymmetric TFD state $|\beta(0, t_R)\rangle$. The fact that the fermions $\psi^j$ and $\psi^{j+N/2}$ are coupled (unlike in the TFD where there is no coupling between $\psi^j_L$ and $\psi^j_R$) should be important at higher orders in $1/N$, but it doesn't matter at leading order.

The above operator size has the same problem as the time-shifted TFD, in that the prefactor becomes smaller as we increase $t_R$. When $t_R \gg \beta$, the  growth will have a very small prefactor, but we want to point out that the operator size still correctly computes the size of monomials, so it seems reasonable to believe that our derivation is still valid. It is only when $1/N$ corrections build up that Equation \ref{BstateSize} will start failing. At late times when the off-diagonal correlation functions die out and the black hole equilibrates, our formula for operator size will no longer capture the leading $1/N$ behavior. 

Instead, we follow a similar argument as to the late-time TFD and use the mirror operatosr $\tilde{\psi}^j(t_R)$ to create the operator size
\begin{equation}\label{MicrostateSize}
\hat{S}_{|B_s(\beta; t_R)\rangle} = \sum_{j=1}^N \frac{\langle i \tilde{\psi}^j(t_R) \psi^j(t_R) \rangle_\beta -i \tilde{\psi}^j(t_R) \psi^j(t_R)}{2\langle i  \tilde{\psi}^j(t_R) \psi^j(t_R) \rangle_\beta}
\end{equation}
At this point we want to point out a subtlety that we left unaddressed earlier. Why should we have used the mirror operator $\tilde{\psi}^j(t_R)$ and not some other mirror operators $\tilde{\psi}(t_R')$? In our derivations that didn't rely on mirror operators, we chose to build the ansatz from Majorana fermions because they were simple operators, and we wanted to have an expression that remained invariant in the large-$N$ limit. Furthermore, we wanted an expression that correctly counts the size of single-sided monomials on both sides, and that forced us to use $\psi^j_L(t_L), \psi^j_R(t_R)$ instead of Majorana fermions at some different times. However, when it comes to mirror operators, it's not immediately clear that the size of monomials in $\tilde{\psi}^j(t_R)$ should be the ``naive size"\footnote{The mirror operators do satisfy an analogue of the KMS condition \ref{KMS}, and thus it is reasonable to assume that their action on the microstate will be relatively simple. Any mirror operators $\tilde{\psi}(t'_R)$ won't satisfy as simple a relation, so we'd expect them to have a larger operator size. This seems to suggest that the size of mirror operator monomials should just be the degree of the monomial. This statement should be rigorous in the high temperature limit where the KMS conditions simplify (and we can directly map monomials in $\tilde{\psi}(t_R)$ to monomials in $\psi(t_R)$), but we can't prove it in higher generality.}, and furthermore we're using $N$-dependent operators from the start. 

Any choice of mirror operators $\tilde{\psi}(t_R')$ will give an operator size that correctly counts monomials of $\psi(t_R)$, but different such choices correspond to the presence of different ``shortcuts" which give different operator sizes. However, for a typical black hole microstate, we expect that we'll have the minimum number of such shortcuts (i.e. we have no ``extraneous" relations that are satisfied by the microstate), and so it is reasonable that we should pick the fastest growing operator size. As we have seen from Equation \ref{TimeShiftedSize}, this corresponds to a choice of $\tilde{\psi}(t_R)$, and this picks out the operator size of Equation \ref{MicrostateSize}.

As a final note, we want to point out that $\hat{S}_{|B_s(\beta)\rangle}$ is the same operator that was used in \cite{MaldacenaKourkoulou} to reveal part of the black hole's interior. Similarly, the late-time operator in Equation \ref{MicrostateSize} was used in \cite{InteriorTypicalMicrostate} to achieve the same result in typical black hole microstate; and the TFD size $\hat{S}_{|\beta\rangle}$ is the operator that is used to make the AdS$_2$ wormhole traversable \cite{GJW,DivingTraversable}. 

This persistent connection between operator size and black hole traversability was noted in \cite{NearHorizon}, where the size operator $\hat{S}_{|\beta\rangle}$ was a ``key ingredient" in creating a global time-translation symmetry $\hat{E}$ that can move the horizon of the AdS$_2$ wormhole. For a bulk particle, time evolution with $\hat{E}$ amounts to a time-advance that allows it to cross the AdS$_2$ wormhole. From the perspective of a highly boosted particle, $\hat{S}_{|\beta\rangle}$ and $\hat{E}$ are one and the same, and thus time evolution with $\hat{S}_{|\beta\rangle}$ can be used to traverse the wormhole. In Section \ref{CFTSection}, we will find that a similar story holds for 2-dimensional CFTs. 

This section can be safely skipped at a first read, as in Section \ref{CFTSection} we will mostly do ``phenomenological" work that tries to guess the operator $\hat{S}_{|\Psi\rangle}$ from some basic principles, rather than trying to use the kind of monomial-counting we use in this section. 

\section{Operator Size for Free Fields}\label{FreeFieldSize}

In this section, we will apply our construction of $S_{|\Psi\rangle}(\mathcal{O})$ to formulate operator size for free field theories. While the size growth is trivial for free theories, this may be a good warmup for more complicated systems, or perhaps a starting point for a perturbative expansion. As in Section \ref{SYKModel}, our usual method will be to start from Equation \ref{InnerProductForm} and attempt to guess $\hat{S}_{|\Psi\rangle}$ by demanding that it correctly counts the size of monomials. 

The upshot of this section is that the vacuum operator size $\hat{S}_{|\Omega\rangle}$ in free field theories is given by the number operator $n_\phi$. At finite temperature, one instead finds that $\hat{S}_{|\beta\rangle}$ counts the number of ``Kruskal particles" $\Phi_1, \Phi_2$ that have $|\beta\rangle$ as their vacuum. This number operator can be expressed by a bi-local integral over $\phi$ which couples the left and right side of the TFD, mirroring the structure of Equation \ref{ThermalSize3}.

\subsection{A Warmup: the Harmonic Oscillator}

Before we proceed with field theories proper, let us first try to construct $\hat{S}_{|\Psi\rangle}$ for states of the harmonic oscillator. As a starter, let us try to compute the size operator $\hat{S}_{|0\rangle}$ for the ground state $|0\rangle$. Consider an operator $\mathcal{O}$ acting on $|0\rangle$ to create a state
\begin{equation}\label{Ostate}
\mathcal{O}|0\rangle = \sum_n c_n |n\rangle
\end{equation}
First thing to note is that any monomial in $a, a^\dagger$ acting on $|0\rangle$ will produce a state $|k\rangle$, it will never produce a superposition of number eigenstates. Thus, if we have a sum of monomials that produces $\sum c_n|n\rangle$, we can't hope for any cancellations between these monomials. If $\mathcal{O}|0\rangle$ was a number eigenstate $|n\rangle$, then the smallest monomial that can create $\mathcal{O}|0\rangle$ is $(a^\dagger)^n$ (up to a prefactor). So generally, the smallest operator that yields \ref{Ostate} is
\begin{equation}
\sum c_n \frac{(a^\dagger)^n}{\sqrt{n!}}
\end{equation}
and we can easily compute the size 
\begin{equation}
\sum |c_n|^2 n
\end{equation}
This is simply the average number of $\mathcal{O}|0\rangle$, and thus we have obtained the result
\begin{equation}
\hat{S}_{|0\rangle} = a^\dagger a
\end{equation}
The size operator for the ground state of the harmonic oscillator is simply the number operator. While this formula was easy to obtain, it is very difficult to do the same for general excited states. In fact, we expect that generally, $\hat{S}_{|\Psi\rangle}$ will not be any simple polynomial of $a, a^\dagger$.

Let's consider the case where $|\Psi\rangle = |n\rangle$. A similar argument to the one above shows
\begin{equation}
\hat{S}_{|n\rangle} = \sum_{m=0}^{\infty} |m-n| \, |m\rangle\langle m|
\end{equation}
By inspection, one can see that this isn't any simple function of $a, a^\dagger$. If we start taking general linear combinations $\sum c_n |n\rangle$ the situation becomes even more difficult to handle, and it is generally impossible to find an explicit formula for $\hat{S}_{|\Psi\rangle}$. Generally, as we saw in Section \ref{SYKModel}, it is easier to construct operator sizes when we have factorizing 2-point functions and vanishing 1-point functions for the fundamental operators. 

Generic excited states won't satisfy these properties, and we can't create a simple size operator for them. However, it is possible to do so for a state of special interest, the thermofield double
\begin{equation}
|\beta\rangle = \sum_{n=0}^{\infty} e^{-\beta\omega/2} |n\rangle_L |n\rangle_R
\end{equation}
To construct its size operator $\hat{S}_\beta$, first recall that we need $\hat{S}_\beta$ to be Hermitian, positive-definite and to annihilate $|\beta\rangle$. This means that we should expect it to be of the form $\sum A_i^\dagger A_i$ where $A_i$ are annihilators of $|\beta\rangle$. The thermofield double satisfies the equations
\begin{equation}\label{TFDconstraints}
(a_L - e^{-\beta\omega/2}a^\dagger_R)|\beta\rangle = (a_R - e^{-\beta\omega/2}a^\dagger_L)|\beta\rangle =  0
\end{equation}
so this gives some candidate building blocks for $\hat{S}_\beta$. By rescaling these annihilators, we get operators
\begin{equation}\label{Bogoliubov}
b_1 = \frac{a_R - e^{-\beta\omega/2}a^\dagger_L}{\sqrt{1 + e^{-\beta\omega}}}, \quad b_2 = \frac{a_L - e^{-\beta\omega/2}a^\dagger_R}{\sqrt{1 + e^{-\beta\omega}}}
\end{equation}
that satisfy the canonical commutation relations $[b_i, b_j] = [b^\dagger_i, b^\dagger_j] = 0$ and $[b_i, b^\dagger_i] = 1$. This is simply a Bogoliubov transformation that expressed the vacuum as the ground state of the operators $b_1, b_2$. Thus, we obtain a natural candidate for the operator size
\begin{equation}
\hat{S}_{|\beta\rangle} = b^\dagger_1 b_1 + b^\dagger_2 b_2
\end{equation}
There is one thing that we must be careful about at this point: while we have obtained a viable size operator, it's not immediately clear that it is the size operator that corresponds to the fundamental fields $a_R, a^\dagger_R, a_L, a^\dagger_L$. As evidence that this is the case, we compute the size of the monomials $a_R^n, (a^\dagger_R)^n$. Since all annihilators of $|\beta\rangle$ involve left and right modes of equal size, we can't use them to shorten a purely right-sided monomial. Thus, we need these monomials to have size $n$. This is easy to obtain by rewriting the $a$'s in terms of the $b$'s; for example
\begin{equation}
a_R = \frac{\sqrt{1 + e^{-\beta\omega}}}{1-e^{-\beta\omega}}(b_1 + e^{-\beta\omega/2}b_2^\dagger)
\end{equation}
The $b_1$ term annihilates $|\beta\rangle$, so it is easy to see that
\begin{equation}
(a_R)^n|\beta\rangle \sim (b_2^\dagger)^n|\beta\rangle
\end{equation}
up to a numerical prefactor. Thus, it immediately follows that the size of $a_R^n$ is $n$, and an identical argument holds for $(a_R^\dagger)^n$. Thus, the $\hat{S}_{|\beta\rangle}$ operator we used appears to be the correct one.

In principle, we still don't know that the above expression is unique. The sizes of the operators $(a_R^\dagger)^n, a_R^n$ aren't sufficient to uniquely determine a size operator. We'd also need to compute products that involve both $a$'s and $a^\dagger$'s, and also mixed left-right products. It is only because we used a quadratic ansatz that we could uniquely determine the form of $\hat{S}_{|\beta\rangle}$, but it's not clear why $\hat{S}_{|\beta\rangle}$ shouldn't be a complicated series of $b$'s instead. 

What is clear is that the $\hat{S}_{|\beta\rangle}$ we wrote is the correct size operator if we take the $b$'s to be the fundamental fields. Since $b$'s and $a$'s are related by a linear transformation and they both satisfy canonical commutation relations, it seems plausible that their respective operators $\hat{S}_{|\beta\rangle}$ are equivalent, but we don't yet have mathematical proof that this is the case. 

\subsection{Free Fields}

Moving on from the harmonic oscillator, let us consider the case of a free scalar field $\phi$ on $D$-dimensional spacetime. We can decompose $\phi$ in terms of creation and annihilation operators $a_p, a^\dagger_p$ as

\begin{equation}\label{SpatialFourier}
\phi(x,t) = \int \frac{d^{D-1}p}{(2\pi)^{D-1}} \frac{1}{2\omega(p)}\big(e^{ipx-i\omega(p) t} a(p) + e^{-ipx+i\omega(p)t} a^\dagger(p)\big)
\end{equation}
where $\omega(p) = \sqrt{p^2 + m^2}$ and the creation/annihilation operators satisfy the canonical commutation relations
\begin{equation}
[a(p), a^\dagger(p')] = 2\omega(p)(2\pi)^{D-1} \delta^{D-1}(p-p')
\end{equation}
The field's Hilbert space decomposes into a direct sum of harmonic oscillators, so we can write an ansatz for the vacuum operator size as
\begin{equation}
\hat{S}_{|\Omega\rangle} = \int \frac{d^{D-1}p}{(2\pi)^{D-1} 2\omega(p)} a^\dagger(p) a(p) f(p)
\end{equation}
where $f(p)$ is some function of the spatial momentum $p$. Due to the translational symmetry of the vacuum, we expect that any reasonable basis of fundamental operators will be translationally invariant. Suppose for example that we have a smeared version of $\phi(x)$ in our basis of fundamental operators, let's call it $V(x)$. Then, any integral
\begin{equation}
\int d^{D-1}x V(x) e^{-ikx}
\end{equation}
will also be a fundamental operator. By using different values of $k$, we can project out all the spatial Fourier components $\phi_k$. However, this isn't a complete basis; we also need to include the canonical momentum $\pi(x)$ into our definition of basis operators. By taking linear combinations of $\phi_k$ and $\pi_k$ we obtain the creation and annihilation operators $a_k, a^\dagger_k$. Thus, in our ansatz we have $f(p) = 1$ and we write
\begin{equation}
\hat{S}_{|\Omega\rangle} = \int \frac{d^{D-1}p}{(2\pi)^{D-1} 2\omega(p)} a^\dagger_p a_p = n_{\phi}
\end{equation}
so the operator size simply counts the number of $\phi$ particles. The operator size is equal to the total number of $\phi$ particles.

Using canonical commutation relations, it is easy to check that $S_{|\Omega\rangle}(\phi(x,0)) = 1$, but for higher $n$ we have $S_{|\Omega\rangle}(\phi(x,0)^n) < n$. This isn't too surprising as $\phi$ both creates and destroys particles, so when it is applied $n$ times we will have less than $n$ particles. This manifests as contractions of the $\phi$'s on the same side of $\hat{S}_{\Omega}$ that reduce the ratio in \ref{InnerProductForm}. If we want to create an $n$-particle state, a natural choice is to use the normal ordered operator $:\phi^n:$ which subtracts all such contractions, and then Wick's theorem immediately yields $S_{|\Omega\rangle}(:\phi(x,0)^n:) = n$.

The above definition of operator size was written in momentum space, but ideally we would want to have a position space definition as well. To do so, we start by separating $\phi$ into its positive and negative frequency parts $\phi_+(x,t), \phi_-(x,t)$. The positive frequency part only contains creation operators, while the negative frequency part only contains annihilation operators. The number operator can then be written as
\begin{equation}\label{PositionSpace}
n_{\phi} = \int d^{D-1}x \phi_+(x,0) (\sqrt{-\nabla_x^2 + m^2})\phi_-(x,0)
\end{equation}
The square root $\sqrt{-\nabla_x^2 + m^2}$ is a well-defined operator (acting on the space of fields) since $-\nabla_x^2 + m^2 = p_x^2 + m^2$ is positive definite, and such operators have a unique Hermitian, positive-definite square root. One thing to note is that it isn't a local operator, but it is a short-hand for a bilocal integral with a kernel $K_m(x,y)$. In principle, this can be computed explicitly from the Fourier transform, but here we just want to note that the non-locality involved is of order $1/m$. This is a natural restriction, since massive particles in field theory are unable to be localized more precisely than $\Delta x \sim 1/m$. For all practical purposes, switching to the momentum space representation will be simpler.

Now that we have a size operator for the vacuum, we can follow a similar method to what we did in the harmonic oscillator to obtain the size operator for the thermofield double
\begin{equation}\label{SizeMomentumAppendix}
\hat{S}_{|\beta\rangle} = \int \frac{d^{D-1}p}{(2\pi)^{D-1} 2\omega(p)} \big(b_{1,p}^\dagger b_{1,p} + b_{2,p}^\dagger b_{2,p}\big)
\end{equation}
where $b_{1,p}, b_{2,p}$ are defined by a Bogoliubov transformation of $a_p, a^\dagger_p$ just as in Equation \ref{Bogoliubov}. A position-dependent expression will be difficult to write in general, as the fields $\Phi_1, \Phi_2$ created from $b_1, b_2$ are not simple linear combinations of $\phi_L, \phi_R$. However, it will be of the form 
\begin{equation}\label{SizePositionAppendix}
\hat{S}_{|\beta\rangle} = \sum_{a,b;i,j}\int d^{D-1}x d^{D-1}y \, \phi^{a}_i(x,0) K_{ij}^{ab}(x,y) \phi^b_j(y,0)
\end{equation}
where $i,j$ run over $L,R$ and $a,b$ run over the positive/negative frequency parts. The kernel $K^{ab}_{ij}(x,y)$ can be in principle obtained by expanding out the $b$'s and the $\phi$'s in Equations \ref{SizeMomentumAppendix}, \ref{SizePositionAppendix} in terms of $a$'s. Matching terms on both sides will give the Fourier transform of the kernel. This expression is very similar to our results in Section \ref{SYKModel}, and once again we see a left-right coupling gives a similar kind of double-trace deformation as the one used in \cite{GJW} to render a wormhole traversable.

\section{Operator Size in Holographic Field Theories}\label{CFTSection}

While in SYK and for free theories it was easy to find a set of fundamental operators, and thus construct an operator size, it's not as easy to do so for an interacting theory. A major problem is that local fields must be smeared both in space and in time in order to give a well-defined operator. This wasn't a problem for a free theory, where the dispersion relation $\omega = 
\sqrt{k^2 + m^2}$ ensures that only a spatial smear is necessary to isolate the creation/annihilation operators (which are well-defined). For an interacting theory, each local field $\mathcal{O}_k(t) = \int d^{D-1}x \mathcal{O}(x,t)$ is a sum of infinitely many operators 
\begin{equation}
\mathcal{O}_k(t) = \int d\omega \, \mathcal{O}_{k,\omega} e^{-i\omega t}
\end{equation}
A smear over time is necessary in order to suppress the high energy modes and give a well-defined operator.

But while we could use spatial translational symmetry to ensure that any spatial smearing was allowed, and thus obtain the creation/annihilation operators, we cannot apply the same argument in a temporal direction. If we were to allow all operators $\mathcal{O}_{k,\omega}$ to be fundamental operators, then we would get a trivial operator size. One would have to choose a (non-unique) smearing function to define a non-trivial operator size, and as we will soon see any such attempt won't give a particularly meaningful definition\footnote{One could also try to write an operator size in terms of the ``microscopic" fields of the holographic gauge theory. We briefly discuss such approaches in the Discussion section.}.

What we will instead do is follow a ``phenomenological" approach, where our starting point is the intuition that OTOCs measure the size of an operator $\mathcal{O}(t)$ in some sense. One may ask at this point, why bother defining an operator size at all and not just use the OTOC? 

One reason is that an OTOC $\langle V_1 W_2 V_3 W_4\rangle$ is dependent on the exact configuration of the operators, and thus it's not clear which OTOC we should use to measure operator size. For states like the thermofield double $|\beta\rangle$, there doesn't seem to be much of a room for error, each OTOC will be of the form
\begin{equation}
\frac{\Delta_V \Delta_W}{c \sin(\frac{2\pi\tau_V}{\beta})} e^{\frac{2 \pi t}{\beta}}f(x)
\end{equation}
where $c$ is the central charge and $\tau_V$ is a Euclidean time evolution used to make the energy of $V$ finite, so it seems easy to take out the spatial dependence and declare that the size of $V(-t-i\tau_V)$ will be
\begin{equation}
\frac{\Delta_V}{\sin(\frac{2\pi\tau_V}{\beta})}e^{\frac{2\pi t}{\beta}}
\end{equation}
However, there are some subtleties if we consider the operator size for a state below the black hole threshold. In \cite{ScramblingPhases}, the authors found an oscillating OTOC for the AdS$_3$ vacuum that's proportional to
\begin{equation}\label{OscillatingOTOC}
\frac{\Delta_V }{c \sin(\frac{\tau_V}{l_{AdS}})} \sin^2(\frac{t-x}{2})
\end{equation} 
for $t > x$. This suggests an oscillating behavior for the operator size, but what if we integrated over all $x$? Then it appears that we could get a constant size, so there is some ambiguity involved. We need some way to extract the operator size from an appropriate ``average OTOC".

A problem with using a construction of the form

\begin{equation}
S_{|\Psi\rangle}(\mathcal{O}) = \frac{\sum_V c_V \langle [\mathcal{O}, V]^2 \rangle_{|\Psi\rangle}}{\langle \mathcal{O}^\dagger \mathcal{O}\rangle_{|\Psi\rangle}}
\end{equation}
is that the connected contribution of each OTOC is $O(1/c)$. In order to get an $O(1)$ size, we would need to use an $O(c)$ number of fields, as we did in SYK. Unfortunately, in holographic theories we only have an $O(1)$ number of light primaries. Including heavy primaries would be in tension with our attempt to build an operator size at leading order in $1/c$, so we would either need to use an $O(c)$ number of descendants, or to choose coefficients $c_V \sim O(c)$. The problem with both of these attempts is that the identity exchange (i.e. the disconnected part of the OTOCs) would get enhanced by an $O(c)$ factor and thus all light primaries $\mathcal{O}(t)$ would have an $O(c)$ size regardless of $t$\footnote{This is the same problem we would encounter if we tried to build an operator size in the form of Equation \ref{SizePositionAppendix}. Trying to treat light primaries as generalized free fields and building an operator size in this way yields a nearly-constant size without any significant growth.}. In order to avoid this issue, we would want to remove the identity exchange from the OTOCs, and only include the contributions from the stress tensor and other primaries. However, any contribution from primaries that doesn't get enhanced by relative boosts could only contribute an $O(1)$ amount to the operator size, so we will focus on the growing part of the OTOCs. By ``growing OTOCs" we don't only mean thermal OTOCs, but any OTOC that can be enhanced by the appropriate kinematics. In the vacuum for example, OTOCs can grow large in the limit of large relative boost.

At the level of operators, what we will want to do is to construct $\hat{S}_{|\Psi\rangle}$ from a linear combination of operators that appear in the OPEs responsible for growing OTOCs. For the vacuum  $|\Omega\rangle$ (which is the first state we will consider), this suggests using integrals of the stress energy tensor\footnote{If we were to consider an OTOC taken in a heavy state $V_H|\Omega\rangle$, then the exchanged mode is the ``dressed" stress-energy tensor, referred to as a ``scramblon" in \cite{ScramblingPhases}. The situation can also get complicated for double-sided states. In a TFD for example, it isn't clear how we should treat the ``OPE" of two operators $V_L V_R$ even though any OTOC $\langle W_R V_L V_R W_R \rangle_\beta$ grows. In the bulk, one could consider the geodesic operator that connects the two sides, but it's unclear how to interpret this as an operator statement on the boundary side, given that the operators don't belong to the same CFT.}\cite{ShockwaveOPE}. This is analogous to the structure of the size operator in SYK, which in \cite{NearHorizon} was written in terms of symmetry generators that were linear in the reparametrization modes $\epsilon_{l/r}(\tilde{u}) = t_{l/r}(\tilde{u}) - \tilde{u}$ and their derivatives. In the CFT vacuum, the exchange of the stress-energy tensor plays the same role as the reparametrization mode\footnote{Recent work on reparametrization theories at 2 dimensions suggests \cite{AdS3Reparametrizations,ChaosEFT,CFTQuantumChaos} that perhaps we could write the operator size in terms of reparametrization modes as in SYK. We leave this approach for future work.}, so we use an ansatz for the operator size (at leading order in $1/c$)
\begin{equation}\label{StressTensorAnsatz}
\hat{S}_{|\Omega\rangle} = \int dx dt\, f(x,t) T(x,t) + \bar{f}(x,t) \bar{T}(x,t)
\end{equation}
At higher orders in $1/c$, we should include higher trace contributions built from the stress-energy tensor. However, these shouldn't be present at leading order. The reason is the same combinatorial argument that ensures that only quadratic terms in $\psi$ appeared in \ref{ThermalSize3}, and it boils down to the fact that operator size must be additive at the probe limit (i.e. when we don't include backreaction),
\begin{equation}
\hat{S}_{|\Omega\rangle}(\mathcal{O}_1\mathcal{O}_2) = \hat{S}_{|\Omega\rangle}(\mathcal{O}_1) + \hat{S}_{|\Omega\rangle}(\mathcal{O}_2) + O(1/c)
\end{equation}
Before we go on to find an explicit form for $\hat{S}_{|\Omega\rangle}$, let us momentarily try to understand the above ansatz (namely, forming the operator size from growing parts of the OPE) from a bulk perspective. We take the global AdS$_3$ vacuum as an example, and consider an OTOC $\langle V^\dagger(0,-t) W^\dagger(0, 0) V(0, -t) W(0,0)\rangle$, where $0 < t < 2\pi l_{AdS}$. We can write
\begin{equation}\label{VacuumReflection}
W(0,0)|\Omega\rangle = W(-\pi l_{AdS}, \pi)|\Omega\rangle
\end{equation}
This follows from the fact that for any regularized operator $W(\phi, t-i\epsilon)$ we have
\begin{equation}
\langle W^\dagger(\phi, t+i\epsilon)W(\phi, t-i\epsilon)\rangle = \langle W^\dagger(\phi, t+i\epsilon)W(\phi+\pi, t-\pi l_{AdS}-i\epsilon)\rangle
\end{equation}
By the Cauchy-Schwarz inequality, this implies Equation \ref{VacuumReflection}. Since $W(-\pi l_{AdS}, \pi)$ is spacelike to $V(0, -t)$, we can commute it across $V$ to write the OTOC as 
\begin{equation}
\langle V^\dagger(0,-t) W^\dagger(0, 0) W(-\pi l_{AdS}, \pi)V(0, -t) \rangle
\end{equation}
Suppose that we smear $V, W$ (or equivalently apply a Euclidean time evolution). Then, in the geodesic approximation, this 4-point function measures the length of the almost light-like geodesic that connects $(0,0)$ to $(\pi l_{AdS}, \pi)$ in the presence of a shockwave generated by $V(0,-t)$. From this perspective, we expect a form of the operator size 
\begin{equation}\label{GeodesicAnsatz}
\hat{S}_{|\Omega\rangle} = \int dx_1 dx_2 dt_1 dt_2\, F(x_1, t_1; x_2, t_2) \delta \hat{\mathcal{L}}(x_1, t_1; x_2, t_2)
\end{equation}
where $\delta \hat{\mathcal{L}}(x_1, t_1; x_2, t_2)$ is a bulk operator that measures the length change of the geodesic going from $(x_1, t_1)$ to $(x_2, t_2)$. By the first law of entanglement, the length of a spacelike geodesic connecting points $(u, v)$ on the boundary is given by the modular Hamiltonian on the same interval, which is given by an integral of the boundary stress tensor $T^{\mu\nu}$, so this expression is compatible with \ref{StressTensorAnsatz}. 

Equation \ref{GeodesicAnsatz} can be written as a limit of a double trace deformation, since in the geodesic approximation we have the OPE
\begin{equation}
\frac{\mathcal{O}^\dagger(x)\mathcal{O}(y)}{\langle \mathcal{O}^\dagger(x) \mathcal{O}(y)\rangle} = 1 - \Delta \delta\hat{\mathcal{L}}(x,y)
\end{equation}
This turns Equation \ref{GeodesicAnsatz} into a more similar form to Equations \ref{ThermalSize3}, \ref{SizePositionAppendix}. 

In the CFT vacuum, the first law of entanglement ensured that this bilocal expression is equivalent to a local integral of the stress-energy tensor. But when we try to formulate operator size for double-sided theories, it is clear that these ansatzes cannot be equivalent. Of course, there's the obvious issue that an operator that couples the two sides cannot be equivalent to an integral of single-sided operators. But even if we were to consider a more general ansatz on the CFT side, there's the problem that a left-right geodesic would have to be equivalent to some sort of OPE between operators on different CFTs. This may be sensible from the bulk perspective, where we can define state-depenent geodesic operators $\delta \hat{\mathcal{L}}$ (which are valid within a subspace of bulk states), but it's not clear what this construction would entail from the boundary side. However, we will see that there are certain cases (namely, AdS-Rindler) where the connection is easier to make.

\subsection{AdS$_3$ Vacuum}\label{AdSSection}

In AdS$_3$, we have a reflection symmetry which exchanges the left-moving and right-moving stress-energy tensors $T, \bar{T}$. Thus, in our ansatz \ref{StressTensorAnsatz} we must have $f(x,t) = \bar{f}(x,t)$. Furthermore, rotational symmetry implies that $f(x,t) = f(t)$, and thus 
\begin{equation}
\hat{S}_{|\Omega\rangle} = \int dt f(t) \int dx \, T_{00}(x,t)
\end{equation}
The $dx$ integral simply gives us the CFT Hamiltonian $\hat{H}_{CFT}$ which is time-independent, and thus up to a proportionality constant we have
\begin{equation}
\hat{S}_{|\Omega\rangle} = l_{AdS} \hat{H}_{CFT}
\end{equation}
where we introduced a factor of $l_{AdS}$ to make the expression unitless. On grounds of symmetry and using an ansatz built from the stress-energy tensor, we have found that the operator size is uniquely determined. One may wonder why we obtained an energy here, while operator size was given by a particle number operator for free field theories in Section \ref{FreeFieldSize}.

This may be simply a result of working at strong-coupling, where a notion of particle number is ill-suited. At large-$N$, one could define a particle number operator from the modes of single-trace operators, and from a bulk perspective this simply counts the number of bulk particles. Such an operator size would be unsuitable for a gravitational theory, as it predicts the same size for all particles regardless of energy or position. In fact, if we follow the results of Section \ref{FreeFieldSize} to generalize this definition to TFD states, we would obtain no appreciable growth for operator size.

If we accept that the operator size is proportional to the $\hat{H}_{CFT}$, then we see that despite the oscillating behavior of the OTOC \ref{OscillatingOTOC}, the size is constant. We want to point out that the same is true for the complexity increase of the state $\mathcal{O}|\Omega\rangle$ compared to the vacuum state $|\Omega\rangle$, as measured by the Complexity = Volume conjecture. 

In fact, at linearized order the complexity increase is directly proportional to the operator size. To see this, it is convenient to use a kinematic space formula (ignoring some numerical prefactors)
\begin{equation}
\delta V = \int dx dy \frac{\partial^2 \mathcal{L}(x,0; y, 0)}{\partial x \partial y} \delta \mathcal{L}(x,0 ; y,0)
\end{equation}
The factor $\frac{\partial^2 \mathcal{L}(x,0; y, 0)}{\partial x \partial y}$ is the Crofton form, which provides a natural measure for the space of geodesics living on the $t = 0$ slice. For a static slice, one can compute the length of a bulk curve $\Gamma$ living on the slice by looking at the Crofton measure of geodesics that intersect $\Gamma$. Similarly, one can compute the volume of a bulk region $A$ by integrating the lengths of the chords $\gamma(x,y)\cap A$ with the Crofton form (where $\gamma(x,y)$ is a geodesic connecting $(x,0)$ to $(y,0)$). 

If we perturb the geometry at linear order, the intersection numbers of the geodesics with any bulk curve/region don't change, and the above formulas still hold as long as we account for the change in the Crofton form and the chord lengths\footnote{We thank Bartlomiej Czech for suggesting this argument to us.}. When computing the volume of the entire static slice, the chord length is simply the perturbed length $\mathcal{L}(x,0;y,0) + \delta \mathcal{L}(x,0;y,0)$, so we write
\begin{equation}
\delta V = \int dx dy \frac{\partial^2 \mathcal{L}(x,0; y, 0)}{\partial x \partial y}  \delta\mathcal{L}(x,0 ; y,0)+\frac{\partial^2 \delta\mathcal{L}(x,0; y, 0)}{\partial x \partial y}  \mathcal{L}(x,0 ; y,0)
\end{equation}
We can integrate by parts twice to ensure that no derivatives act on $\delta \mathcal{L}$, and then we have (up to numerical prefactors)
\begin{equation}\label{KinematicVolume}
\delta V = \int dx dy \frac{\partial^2 \mathcal{L}(x,0; y, 0)}{\partial x \partial y}  \delta\mathcal{L}(x,0 ; y,0)
\end{equation}
By the first law of entanglement \cite{GravityDualsModular}\cite{RelativeEntropyBulk}, each $\delta \mathcal{L}$ factor is an integral of $T_{00}$ (with a $1/c = G_N/l_{AdS}$ prefactor) and by rotational symmetry we obtain a multiple of the Hamiltonian. Up to numerical factors, we then have
\begin{equation}
\hat{S}_{|\Omega\rangle} = l_{AdS} \hat{H}_{CFT} \sim \frac{\delta \hat{V}}{l_{AdS} G_N}
\end{equation}
where we promoted the volume change $\delta \hat{V}$ to an operator acting on the subspace of states that are close to the vacuum. We have thus found that for low-energy operators, the size is proportional to the complexity increase, and they are both measured by the energy of an excitation (a similar result related the volume increase to the energy of a scalar field in \cite{NewYorkTime}). This proportionality was suggested in \cite{ComplexityLaws} to hold until backreaction becomes important, but here we find a lack of the suggested oscillatory behavior. This behavior is present in the eternal traversable wormhole of AdS$_2$, but in AdS$_3$ we see that even though the spatial size distribution can vary (in the sense that certain OTOCs are oscillatory), the total size remains constant. In AdS$_2$, the oscillation frequency was related to the ``breathing mode" of the wormhole \cite{ComplexityGeometry}. This was of order $1/l_{AdS}$ in the eternal traversable wormhole \cite{EternalTraversable}, but in general models it can be different. 

We find the situation to be qualitatively different in AdS$_3$ (and in higher dimensions, where our arguments generalize); if any oscillations are present then they are subleading in $1/c$. Note that this difference is present both for operator size, and also for complexity. In Section \ref{ComplexitySection}, we will further discuss the leading-order proportionality between operator size and complexity.

\subsection{AdS-Rindler and the TFD}

\begin{figure}
\begin{center}
\includegraphics[height=6cm]{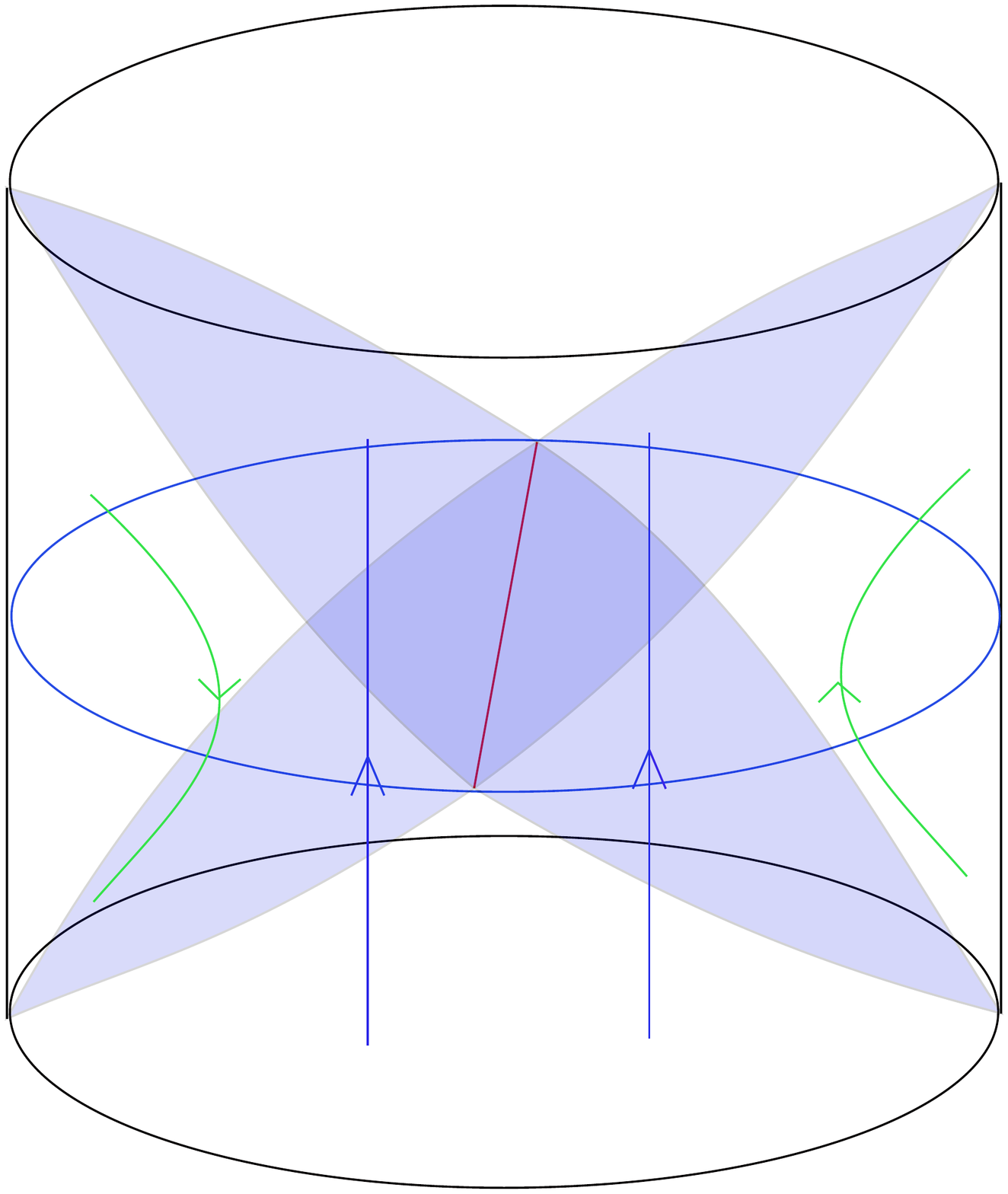}
\caption{The AdS$_3$ vacuum can be expressed in as an entangled state of two line CFTs. In the bulk, this corresponds to using accelerating coordinates which produce the horizons which are shaded blue (the red line denotes the bifurcation surface). The green/blue arrows show the Rindler/global Hamiltonian evolution in the bulk respectively.}\label{fig:AdSRindler}
\end{center}
\end{figure}

We now wish to use our result for the AdS$_3$ vacuum to understand operator size in the simplest ``black hole" state, the AdS-Rindler geometry. The CFT vacuum we considered above can be related to a TFD state by switching to Rindler coordinates \cite{RindlerAdSCFT}\cite{RindlerQuantumGravity}. Just as the Minkowski vacuum can be written as a thermofield double, the CFT vacuum on a cylinder can be expressed as a thermally entangled state of two CFTs living on a line
\begin{equation}
|\Omega\rangle = \sum_E e^{-\pi E l_{AdS}} |E\rangle_L |E\rangle_R
\end{equation}
In terms of the original CFTs, we take the two CFTs to respectively live on the intervals $\phi \in (0, \pi)$ and $\phi \in (\pi, 2\pi)$ of the $t=0$ slice, and they evolve with the modular Hamiltonians $K_L, K_R$ that correspond to these two intervals. A conformal transformation takes the two intervals to a pair of infinite lines, and the modular Hamiltonians become the CFT Hamiltonians $H_L, H_R$. In these $(t,\chi)$ coordinates (where $x$ parametrizes each line), the 2-point functions become thermal
\begin{equation}
\langle \mathcal{O}_R(t, \chi) \mathcal{O}_R(0,0)\rangle = \frac{1}{\big( \cosh(\frac{t}{l_{AdS}}) - \cosh(\frac{\chi}{l_{AdS}}) \big)^{\Delta}}
\end{equation}
\begin{equation}
\langle \mathcal{O}_R(t, \chi) \mathcal{O}_L(0,0)\rangle = \frac{1}{\big( \cosh(\frac{t}{l_{AdS}}) + \cosh(\frac{\chi}{l_{AdS}}) \big)^{\Delta}}
\end{equation}
In the bulk, we have the AdS-Rindler geometry which corresponds to a uniformly accelerating observer. This geometry has an acceleration horizon which separates the causal wedges of the $(0,\pi)$ and $(\pi, 2\pi)$ intervals (see Figure \ref{fig:AdSRindler}). This is most conveniently seen in Kruskal coordinates where the geometry becomes
\begin{equation}\label{AdSRindlerMetric}
ds^2 = l_{AdS}^2 \frac{-4 du dv}{(1+uv)^2} + \frac{(1 - uv)^2}{(1+uv)^2}d\chi^2
\end{equation}
In this coordinate system, the horizons are given by $u = 0, v= 0$, and the boundary is $uv = -1$. On the boundary, we can relate $u,v$ to the asymptotic time $t$ by $u = -1/v = e^{t/l_{AdS}}$. This geometry is identical to the BTZ black hole up to a quotient $\chi = \chi + 2\pi r_s$, so it provides a simple model for operator growth in black holes.

It is clear that we are considering the same state as before, but one cannot immediately conclude that $\hat{S}_{|TFD\rangle} = \hat{S}_{|\Omega\rangle} = l_{AdS} \hat{H}_{CFT}$. There is always the possibility that there's a different basis of fundamental operators in the two cases. However, primaries transform simply under conformal transformations (they only pick up a scaling factor), and thus any conformal transformation that preserves the $t=0$ slice won't change the size of primaries living on it. It seems plausible then that the fundamental operators we implicitly used in our construction transform simply under such conformal transformations, in which case we can use the same operator size\footnote{At the end of the day, we can choose to use the same basis of fundamental operators. Still, since we're working phenomenologically and we didn't have detailed rigorous definition in the first place, it is worth checking that such a choice is plausible.}

\begin{equation}
\hat{S}_{|TFD\rangle} = l_{AdS} \hat{H}_{CFT}
\end{equation}
Here, we wrote $|TFD\rangle$ to emphasize the coordinate system we are using, and that when we compute the size $S_{|TFD\rangle}(\mathcal{O}_R(-t))$ of an operator we are using the Hamiltonian $\hat{H}_R$ rather than the global Hamiltonian $\hat{H}_{CFT}$ to evolve the operator $\mathcal{O}_R$. The Hamiltonian $\hat{H}_R$ doesn't commute with $\hat{H}_{CFT}$, and thus the size growth of $\mathcal{O}_R(-t)$ will be non-trivial. Evolving an excitation with $\hat{H}_R$ increases its boost $\eta$ relative to the frame of the global $t = 0$ slice, and thus the global energy will increase as 
\begin{equation}
E_{global} \sim E_0 e^{\eta} = E_0 e^{t/l_{AdS}}
\end{equation}
To see the connection with OTOCs, let us consider an operator $\mathcal{O}_R(-t)$ that creates an infalling excitation, with $t \gg l_{AdS}$. Then, the bulk excitation will be in the near-horizon region of the geometry, and we can relate its global energy to the Kruskal momenta
\begin{equation}
\hat{H}_{CFT} \sim \frac{P_u + P_v}{l_{AdS}}
\end{equation}
Note that we picked our Kruskal coordinates $u,v$ to be dimensionless, and thus the momenta are dimensionless as well. The shockwave created by $\mathcal{O}_R(-t)$ will cause a delay proportional to
\begin{equation}
\frac{G_N P_v}{l_{AdS}} \sim \frac{\Delta_{\mathcal{O}}}{\sin(\frac{\tau}{l_{AdS}})}e^{t/l_{AdS}}
\end{equation}
where $\tau$ is a Euclidean parameter used to regulate the energy of $\mathcal{O}_R$. Thus, the typical OTOC $\langle \mathcal{O}^\dagger_R(-t) V^\dagger_L V_R \mathcal{O}_R(-t)\rangle$ will indeed be proportional to the size we defined, up to an $x$-dependent function which indicates the spatial profile of the shockwave. Before we proceed to the case of a BTZ black hole, we want to discuss the connection to traversability.

In the SYK model, there was a clear connection between operator size and the double-trace deformation that allows an excitation to traverse a wormhole. In the AdS$_3$ situation, since $\hat{S}_{|TFD\rangle}$ is proportional to the global Hamiltonian, it is clear that a deformation $\epsilon(t) \hat{S}_{|TFD\rangle}$ will allow an excitation to cross the ``wormhole". For example, let us consider
\begin{equation}
e^{i\hat{S}_{|TFD\rangle}\tau} \mathcal{O}_R(-t)e^{-i\hat{S}_{|TFD\rangle}\tau}
\end{equation}
Even though $\mathcal{O}_R(-t)$ remains outside the causal past of the left side no matter how large $t$ is, evolving backwards with the global Hamiltonian for a time
\begin{equation}\label{HamiltonianPush}
\tau \sim -e^{-t/l_{AdS}}
\end{equation}
will move $\mathcal{O}_R(-t)$ into the causal past of the left side, allowing it to cross the horizon. Of course, there's nothing surprising about being able to cross a Rindler horizon using the global Hamiltonian, but we want to point out that our kinematic space Equation \ref{KinematicVolume} can be used to relate this trivial traversability to a double trace deformation. 

The length of a geodesic can be estimated by computing two-point functions \cite{NearHorizon}, and thus we can re-write the kinematic volume formula as
\begin{equation}
\delta \hat{V} \sim \frac{l_{AdS}}{K}\sum_{i, a, b} \int d\chi_1 d\chi_2 \frac{\partial^2 \mathcal{L}_{ab}(\chi_1,0; \chi_2, 0)}{\partial \chi_1 \partial \chi_2}   \frac{\langle \mathcal{O}^i_a(0, \chi_1) \mathcal{O}^i_b(0, \chi_2)\rangle-\mathcal{O}^i_a(0, \chi_1) \mathcal{O}^i_b(0, \chi_2)}{\Delta_i\langle \mathcal{O}^i_a(0, \chi_1) \mathcal{O}^i_b(0, \chi_2)\rangle}
\end{equation}
where $i$ runs over a set of $K$ light single trace operators, and $a,b$ runs over the left and right sides of the geometry. This formula is valid as long as we have a large number of primaries $\mathcal{O}^i$ and we only care about linearized gravity. Due to the relation between the linearized volume variation and the global Hamiltonian, we can thus write

\begin{equation}\label{DoubleTraceHamiltonian}
\hat{H}_{CFT} \sim \frac{1}{G_N l_{AdS} K}\sum_{i, a, b} \int d\chi_1 d\chi_2 \frac{\partial^2 \mathcal{L}_{ab}(\chi_1,0; \chi_2, 0)}{\partial \chi_1 \partial \chi_2}   \frac{\langle \mathcal{O}^i_a(0, \chi_1) \mathcal{O}^i_b(0, \chi_2)\rangle-\mathcal{O}^i_a(0, \chi_1) \mathcal{O}^i_b(0, \chi_2)}{\Delta_i\langle \mathcal{O}^i_a(0, \chi_1) \mathcal{O}^i_b(0, \chi_2)\rangle}
\end{equation}
As long as we work in the regime where gravity stays in the linearized regime, we can replicate the effect of $\hat{H}_{CFT}$ by using a bilocal double-trace operator. If we wish to implement a version of the GJW protocol using above rewriting of the Hamiltonian as a double-trace deformation, there are two conditions we need to obey. The first one is that we stay in the regime of linearized gravity. The second is that the time evolution of a bulk field $\mathcal{O}$ in the interaction picture $e^{i\tau \int dt H_{CFT}(t)}\mathcal{O} e^{-i\tau  \int dt H_{CFT}(t)}$ is dominated by the first order term in the expansion. If high order terms are included, then there is the risk that we will move out of the regime where Equation \ref{DoubleTraceHamiltonian} is valid\footnote{This is related to a point made in \cite{NearHorizon} that two-point functions become bad probes of distance when a large number of insertions is present. }. This places a constraint on $\tau$ that is
\begin{equation}
\frac{|\tau| l_{AdS}}{G_N} \ll 1 
\end{equation}
From Equation \ref{HamiltonianPush}, we see that an excitation $\mathcal{O}_R(-t)$ can cross the horizon as long as
\begin{equation}
t \gtrsim l_{AdS} \log\Big(\frac{1}{K|\tau|}\Big)
\end{equation}
In the regime $K|\tau| \gg \frac{G_N}{l_{AdS}}$, there is a finite window of time where the traversability protocol can be implemented and when backreaction hasn't grown too strong yet. The above result mirrors those of \cite{GJW} for $\beta \sim l_{AdS}$ (by identifying our parameter $|\tau|\frac{l_{AdS}}{G_N}$ with their coupling $g$), and it demonstrates the connection between operator growth and wormhole traversability persists in higher dimensions. Of course, so far we have relied on a global translation symmetry which won't be present for wormholes with a compact horizon, while traversability has been demonstrated in an enormous class of such geometries \cite{GJW}-\cite{RotatingTraversable}.

\subsection{BTZ Geometries}\label{BTZSection}

The BTZ black hole can be obtained from the metric \ref{AdSRindlerMetric} by taking a quotient $\chi \sim \chi + \frac{4\pi^2 l_{AdS}^2}{\beta} = \chi + 2\pi r_s$. For the strongly coupled boundary theory, taking a quotient isn't a straightforward operation; but in the weakly coupled bulk it amounts to including image contributions to every correlation function. Boundary correlation functions can then be computed by taking the extrapolate limit of bulk fields \cite{EternalBlackHoles}. 

How does the operator size fit into this picture? One could hope that for operators of small size, there is a simple relation between their size in AdS-Rindler and their BTZ size. Suppose that we have a collection of fundamental operators $\lbrace \mathcal{O}_i \rbrace$ on AdS-Rindler. Under the AdS-CFT dictionary, these will map to a collection of bulk operators $\lbrace \Phi_i \rbrace$ which are defined on some code subspace of the bulk theory, and they act on the vacuum $|\Omega_{bulk}\rangle$. When backreaction isn't strong, we should be able to equivalently compute the size of an operator in the bulk and the boundary. In other words, if we have
\begin{equation}
\mathcal{O}|\Psi\rangle = \sum_I c_I \prod_j \mathcal{O}_{i_j} |\Psi\rangle
\end{equation}
then there should be a bulk expression for its dual $\Phi$ that is
\begin{equation}
\Phi |\Omega_{bulk}\rangle= \sum_I c_I \prod_j \Phi_{i_j} |\Omega_{bulk}\rangle
\end{equation}
and the two size computations should agree. One thing to note is that even though the bulk is weakly coupled, we can't expect that its fundamental operators $\lbrace \Phi_i \rbrace$ will be spatially uniform. While we can expect spherical symmetry, we also expect the basis to be radially inhomogeneous. For example, in AdS-Rindler we need the operators $\lbrace \Phi_i \rbrace$ to create excitations with a small global energy, so they shouldn't have significant support near the boundary. This inhomogeneity can allow for non-trivial operator growth even for a weakly coupled bulk theory. 

When we take the quotient that maps AdS$_3$-Rindler to a BTZ geometry, the spherically symmetric basis of fundamental operators $\lbrace \Phi_i \rbrace$ will undergo a simple quotient (we can either think of it as such in position basis, or in momentum basis as projecting out operators whose momenta are incompatible with the quotient) to yield a collection of operators $\lbrace \tilde{\Phi}_i \rbrace$. For any bulk excitation $\tilde{\Phi}$ we can expand it in terms of this collection as
\begin{equation}
\tilde{\Phi}|\Omega_{BTZ}\rangle = \sum_I c_I \prod_j \tilde{\Phi}_{i_j} |\Omega_{BTZ}\rangle
\end{equation}
This suggests that $\lbrace \tilde{\Phi}_i \rbrace$ can be used to form a basis of fundamental operators. We can't guarantee it will be the ``correct" one, but it seems like a very reasonable choice. 

For any light excitation $\Phi_{BTZ}$ on the BTZ geometry we can find an excitation $\Phi$ on AdS-Rindler such that $\Phi_{BTZ} = \tilde{\Phi}$ (where the tilde indicates taking the quotient)\footnote{Here, we assume that $\Phi_{BTZ}$ doesn't belong to the twisted sectors of the theory. For a small string length $l_s \gg r_s$ and weak string coupling, the twisted sectors decouple from the untwisted sector, so we will ignore them.}, and the operator size of $\Phi_{BTZ}$ will be given by
\begin{equation}
S_{|BTZ\rangle}(\Phi_{BTZ}) = S_{|TFD\rangle}(\Phi)
\end{equation}
By relating bulk operators to the boundary theory, we obtain an operator size formula for the CFT thermofield doubles $|\beta\rangle$. 

Since the operator size is easy to compute in AdS-Rindler, this gives a simple way to compute operator size for BTZ geometries. The procedure is for any boundary operator $\mathcal{O}$ to figure out its bulk dual, then lift it to AdS-Rindler, and evaluate its global Hamiltonian there. In the near-horizon region of a BTZ black hole, we can directly write (up to a numerical prefactor)
\begin{equation}
\hat{S}_{|BTZ\rangle} = P_u + P_v
\end{equation}
where $P_u, P_v$ are the Kruskal momenta. This is a universal expression, valid for any non-rotating BTZ black hole regardless of temperature. This universality does appear to raise a question however: how does one justify the temperature-dependence of the Lyapunov exponent? The answer is that the time coordinate is different for AdS-Rindler and BTZ. In both geometries, we have $-v = 1/u = e^{2\pi t/\beta}$ on the boundary, and we can treat $u, v$ as being the same in both geometries (based on the way the quotient was taken), but the boundary times $t$ will not be. Expressions of operator size in terms of $u,v$ will thus be universal for all geometries, but the $t,r$-dependent expressions will be different.

Since $P_u + P_v$ generate an upward translation of the horizon, deforming the Hamiltonian with a term $\delta H(t) = \epsilon(t) \hat{S}_{|BTZ\rangle}$ allows a highly boosted particle to cross the horizon. In analogy with AdS-Rindler, it would be nice to have an approximate double-trace expression for $\hat{S}_{|BTZ\rangle}$. The kinematic space of BTZ black holes is more complicated than AdS-Rindler \cite{KinematicComplexity}, and the analogue of Equation \ref{KinematicVolume} for BTZ involves the use of non-minimal geodesics. This makes it difficult to express in terms of 2-point functions, since we cannot independently tune the coefficient of the non-minimal geodesics. We could try to use the 2-point functions of $k$ different operators, but even then we can only tune $k$ non-minimal geodesics, not the infinite number we need. At the end of the day, the best way to generate $P_u + P_v$ in the near-horizon region is the most prosaic one: take a spatially uniform double-trace deformation

\begin{equation}
\int dx \Big(\langle O_L^\dagger(x,0)O_R(x,0)\rangle_\beta - O_L^\dagger(x,0)O_R(x,0) \Big)
\end{equation}
In the limit of large boosts, the expectation value of the above deformation is proportional to the strength of the infalling shockwave. If we drop a shockwave with momentum $P_u$ and another one with momentum $P_v$, then in the linearized gravity limit the time-delay suffered by a geodesic connecting these two shockwaves will be the sum of the time delays. Thus, by virtue of measuring the time-delay, the double-trace deformation measures $P_u + P_v$.

\subsection{Backreaction and Saturation}\label{SectionBackreaction}

We now wish to ask the question: when do $1/c$ corrections to $\hat{S}_{|BTZ\rangle}$ become relevant? There are two issues to consider in this situation. The first one is that backreaction can become strong enough that we leave the regime of linearized gravity. The second is that the method of images can fail in the bulk due to the aforementioned non-linearity, and thus $\hat{S}_{|BTZ\rangle}$ may not ``inherit" the operator size from AdS$_3$-Rindler.

The second consideration is by definition less restrictive than the first one, so let us consider the issue of backreaction for now. For a spatially localized right-infalling particle, backreaction becomes strong when its Kruskal momentum becomes of order
\begin{equation}
P_v \sim \frac{l_{AdS}}{G_N}
\end{equation}
At such a high boost, an infalling particle will cause time delays of order $\Delta v \sim O(1)$ to light-like geodesics passing within an impact parameter of order $l_{AdS}$. The Kruskal momentum of a particle dropped in at time $-t$ is related to the boundary energy $E$ as

\begin{equation}
P_v \sim \beta E e^{2\pi t/\beta}
\end{equation}
Thus, backreaction will become important at a time
\begin{equation}
t_b \sim \frac{\beta}{2\pi} \log(\frac{l_{AdS}}{G_N \beta E})
\end{equation}
If we take the smallest value that the particle can have while being well-localized in time, $E \sim 1/\beta$, then we get an expression that is similar to the scrambling time
\begin{equation}
t_b \sim \frac{\beta}{2\pi} \log(\frac{l_{AdS}}{G_N}) = t_* - \frac{\beta}{2\pi}\log(\frac{r_s}{l_{AdS}})
\end{equation}
For an AdS-scale black hole we indeed obtain the scrambling time, but if $r_s \gg l_{AdS}$ we see that the backreaction time will be somewhat shorter. The difference is most extreme in AdS$_3$-Rindler, where the scrambling time $t_*$ is infinite (due to the infinite entropy of the acceleration horizon), but the backreaction time $t_b$ is finite. 

Now, suppose that instead of a localized particle, we threw in a spherically symmetric null shell with the same energy $E$. Backreaction then becomes important at a Kruskal momentum of order
\begin{equation}
P_v \sim \frac{r_s}{G_N} \sim S_{BH}
\end{equation}
and thus operator size becomes comparable to the black hole entropy before $1/c$ corrections become important. Conversely, for the localized particle the operator size is only $l_{AdS}/G_N$ which is much smaller than $S_{BH}$ for $r_s \gg l_{AdS}$. We see that backreaction is sensitively dependent on the transverse profile of the infalling excitation, and this means that the operator growth is non-universal beyond the probe limit. This suggests that higher $1/c$ corrections will depend on the Kruskal momentum density (in the $\chi$ direction), not just on the Kruskal momenta. 

\begin{figure}
\begin{center}
\includegraphics[height=6cm]{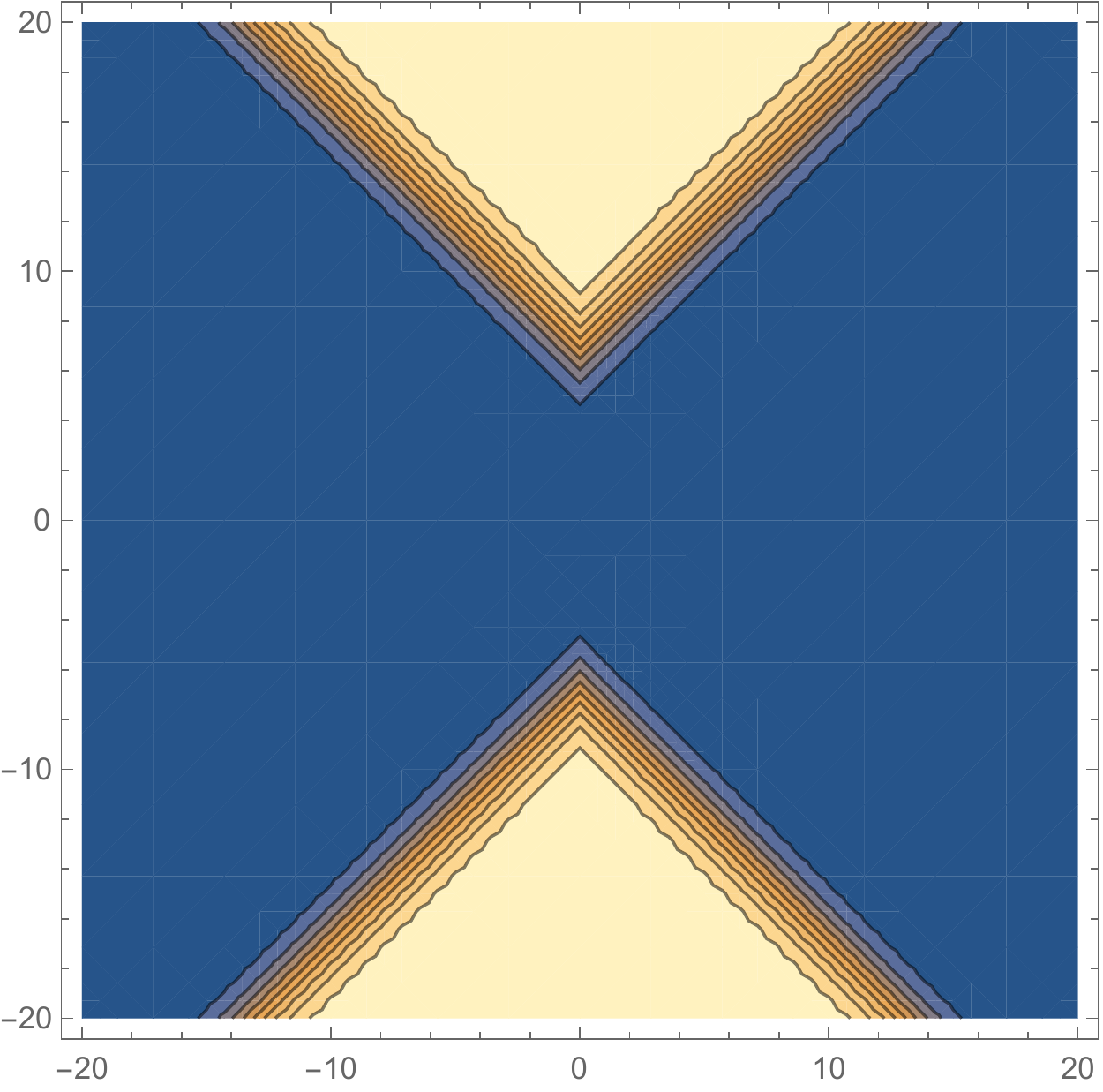}
\caption{The plot shows the typical growth structure of OTOCs $\langle O^\dagger_R(-t,0) V_L(0,x) V_R(0,x) O_R(-t,0) \rangle$. The velocity of this lightcone is the butterfly velocity, which in 2 dimensions is equal to the speed of light.}\label{fig:OTOC}
\end{center}
\end{figure}

This situation is very unlike SYK, where backreaction and scrambling went hand in hand. This difference originates in the fact that SYK is completely non-local, while the CFTs have local interactions. Thus, while the OTOCs in SYK all saturate at the same time, yielding a clear transition from exponential growth to saturation, while in CFT OTOCs involving localized excitations demonstrates a light-cone structure \cite{LocalizedShocks}. If we consider an excitation $\mathcal{O}(-t, 0)$, then within the light-cone
\begin{equation}\label{SaturationLightcone}
v_B |\chi| < t - t_b
\end{equation}
the OTOCs are nearly saturated, while outside the light-cone they are still small and exponentially growing (here, $v_B$ is the butterfly velocity in $\chi$ coordinates). In this regime, the vast majority of OTOC growth comes from the region near the light-cone. If operator size (beyond the probe limit) is still measured by some sort of averaged OTOC, this suggests that there is a transition from exponential growth to linear growth. If we rewrite $\chi = r_s \phi$ (so $\phi \sim \phi + 2\pi$), then the ``size density" in $\phi$ coordinates should be\footnote{By size density we the ratio of an operator's size to the size $\Delta \phi$ of the lightcone \ref{SaturationLightcone}.}
\begin{equation}
\frac{r_s}{l_{AdS}}\frac{l_{AdS}}{G_N}
\end{equation}
This estimate arises from the fact that we expect the operator size to be $O(l_{AdS}/G_N)$ when backreaction becomes important at the AdS-scale\footnote{One may wonder why we don't have linear growth when backreaction is important at some sub-AdS scale. The reason is that the eikonal phase which governs OTOCs has an exponential suppression $e^{-\chi/l_{AdS}}$ at AdS scales, but at sub-AdS scales it is simply a power law. Thus, at AdS scales there's a finite butterfly velocity due to the term $e^{\frac{2\pi t}{\beta} - \chi/l_{AdS}}$ that appears in the OTOC, and the growth comes from the increasing size of the saturated OTOC region. At sub-AdS scales, the power law dependence of the eikonal phase is irrelevant compared to the exponential growth of the center-of-mass energy, and the operator growth doesn't come from the saturation region.}. An AdS-scale impact parameter corresponds to an angular size $\delta \phi \sim l_{AdS}/r_s$, so dividing the two we obtain the above estimate. Assuming linear growth, the operator size should then grow as
\begin{equation}
\frac{r_s}{G_N}\frac{t-t_b}{l_{AdS}}
\end{equation}
where $1/l_{AdS}$ is the butterfly velocity in $\phi$ coordinates. This growth will continue until the light-cone fills the entire boundary after a time $t-t_b \sim l_{AdS}$, and thus we expect the saturated operator size to be
\begin{equation}
S_{|BTZ\rangle}(\mathcal{O}(-t,0)) \sim \frac{r_s}{G_N} \sim S_{BH}
\end{equation}
Note that in the limit $r_s \gg l_{AdS}$, we also have $l_{AdS} \gg \beta$ and thus the above saturation time $t_b + l_{AdS}$ will be larger than the scrambling time which can be written as $t_b + \frac{\beta}{\pi}\log(\frac{l_{AdS}}{\beta})$. Thus, we see that for a local theory, a localized excitation scrambles somewhat slower than a delocalized excitation, though their final size is the same.

The above expectations are based on the general form of OTOCs, but they are sensitively dependent on higher $1/c$ corrections which we can't derive by symmetry like with the leading term. It is difficult to guess a ``natural" bulk operator which exhibits the correct growth pattern for both localized excitations (i.e. an exponential growth $\rightarrow$ linear growth $\rightarrow$ saturation transition) and spherical shells (i.e. an exponential growth $\rightarrow$ saturation transition). Of course, one could use an operator of the form
\begin{equation}
\int d\chi(1 - e^{-\delta \mathcal{L}_{LR}(\chi)})
\end{equation}
where $\delta \mathcal{L}_{LR}(\chi)$ is the length variation of a left-right geodesic that connects the points $(t=0, \chi)$ on the two boundaries. However, this operator is rather artificial and far from unique (we could have used more or less any exponential of the geodesic length). Ideally, some sort of bulk energy or rapidity measurement would be preferable. Perhaps it is possible to create such an operator size built from average null energy operators \cite{ANEC} or the average light cone tilts of \cite{GravityDualBoundaryCausality}, but so far we have been unable to do so. In the next section we show a natural albeit ``experimental" definition of a relative rapidity which exhibits the right growth structure.

\subsection{A Bulk Proposal for the Operator Size of Shockwaves}\label{BulkSection}

\begin{figure}
\begin{center}
\includegraphics[height=6cm]{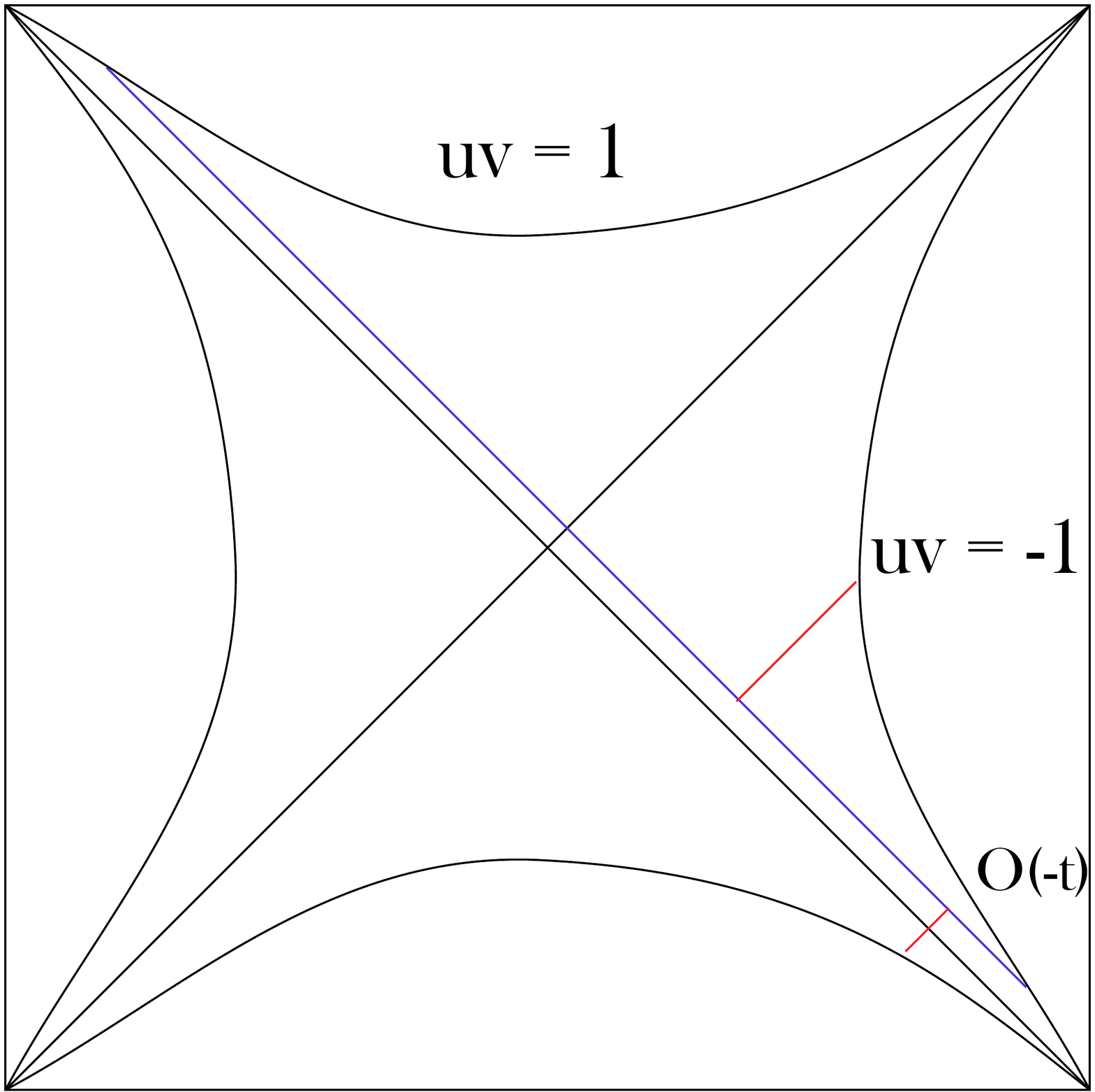}
\caption{As we push $\mathcal{O}(-t)$ to earlier times, it creates an increasingly strong time delay. The pulse that reaches the boundary at $t=0$ must have been emitted earlier than naively expected, thus reducing the relative boost between the pulse and the infalling particle.}\label{fig:shockwaves}
\end{center}
\end{figure}

Consider a lab that hovers at some fixed proper distance $uv \ll 1$ from the horizon, and it emits localized radial pulses with fixed energy $E_0 \gg 1/\beta$ at regular intervals. These pulses are labeled (either by small differences in frequency or some other parameter) so that a boundary observer that receives one of the pulses can tell at what boundary time $t$ it was emitted. Now, suppose that we have a highly boosted infalling particle with momentum $P_v$ that creates a shockwave geometry (see Figure \ref{fig:shockwaves}). As the particle crosses the pulses, it will cause them a delay
\begin{equation}
\Delta u \sim \frac{G_N P_v}{l_{AdS}}e^{-|\Delta\chi|/l_{AdS}}
\end{equation}
where $\Delta \chi$ is the transverse separation of the particle and the pulse. The boundary observer can measure the time-delay of the pulses, and can thus make a measurement of relative rapidity between the particle and the pulse, which is given by the logarithm of the dimensionless center-of-mass energy
\begin{equation}
P_{v,infaller} P_{u,pulse} \sim P_v \beta E_0 e^{2\pi t_0/\beta}
\end{equation}
where $t_0$ is the time when the pulse was emitted. In terms of the $u = e^{-2\pi t/\beta}$ coordinate, the pulse was emitted at $u_0 = 1 + \Delta u$ (since there's a time delay $-\Delta u$ and it arrives at the boundary at $u=1$), so the center-of-mass energy is (up to a factor $\beta E_0$ which is apparatus-dependent)
\begin{equation}\label{RelativeRapidity}
\frac{P_v}{1+\Delta u} \sim \frac{P_v}{1+\frac{G_N P_v}{l_{AdS}}e^{-\Delta \chi/l_{AdS}}}
\end{equation}
We see that no matter how much we increase $P_v$, the relative rapidity will stay bounded above by the quantity
\begin{equation}
\frac{l_{AdS}}{G_N}e^{\Delta \chi/l_{AdS}}
\end{equation}

One caveat in the above analysis was that we assumed that the pulse can be treated as a point particle. As we increase $P_{v,infaller}$ however, $P_{u,pulse}$ decreases and the uncertainty of the pulse $\Delta u_{pulse} \sim 1/P_{u,pulse}$ increases. In order to make the measurement, we must ensure that the entire wavepacket doesn't fall behind the horizon; we will demand that the entire wavepacket (besides some tails) has reached the boundary at time $t=0$, so the emitted time must have been $u_0 = 1 + \Delta u + \Delta u_{pulse}$. By taking $P_{u,pulse} = 1/u_0 \epsilon$, $\Delta u_{pulse} = \epsilon u_0$, we find that the relative rapidity is changed by a factor of $1/\epsilon(1-\epsilon)$. The $1/\epsilon$ is the pulse energy $\beta E_0$ which we divided away in our previous construction, so the only real difference is the $1/(1-\epsilon)$ factor. We can consistently take the limit $\epsilon \rightarrow 0$ after we take $N\rightarrow \infty$ (so $\epsilon$ will always be parametrically $O(1)$ in the $1/N$ expansion), so that we avoid using Planckian pulses and stay within the regime of validity of the eikonal series. The point-particle estimate we made for the above measurement is thus feasible, albeit with a more complicated process.

There is still one problem with our above measurement: it sensitively depends on the transverse separation. If we have an unknown transverse profile for the infalling particle, then we couldn't have made a direct measurement on $P_v$ to measure the relative rapidity. The quantity \ref{RelativeRapidity} isn't a directly measurable quantity then, since it requires knowledge of $P_v$. An estimate of the relative rapidity we can instead measure is
\begin{equation}
\frac{\Delta u}{1+\Delta u}
\end{equation}
Up to a factor $\beta E_0$, this is the eikonal phase $\delta(s, \chi)$ of the scattering between the pulse and the infalling excitation. We can view this as a measure of the gravitational field of the infaller as seen in the frame of the pulse.

Because of spherical symmetry, we are lead to average over all transverse locations of the pulses, which gives the quantity
\begin{equation}\label{ShockwaveSize}
\int \frac{d\chi}{G_N} \frac{\Delta u(\chi)}{1+\Delta u(\chi)}
\end{equation}
where we added a prefactor of $1/G_N$ to fix dimensions and get the correct linearized result\footnote{We use the above formula both for AdS$_3$-Rindler and BTZ black holes. Since the later is obtained from the former by a compactification $\chi \sim \chi + 2\pi r_s$, we see that our proposed formula is compatible with the bulk image method that we used to obtain the leading order $\hat{S}_{|BTZ\rangle}$ operator from $\hat{S}_{|TFD\rangle}$.}. If we want to also consider left infalling particles, we should also consider the analogous expression with $u \rightarrow v$. We want to briefly point out that since $\Delta u(\chi) \sim P_v$, the scaling of this quantity is very similar to the $\hat{E}$ charge constructed in \cite{NearHorizon}, which had an almost identical form (without the $\chi$-dependence). The origin of both effects is similar: backreaction reduces the relative rapidity between the infaller and an appropriately formulated geodesic (which was a left-right geodesic in \cite{NearHorizon}, and a null geodesic in our case). Of course, what we are considering here is a mere toy model, which ideally we would want to formulate more rigorously rather than relying on a fictitious ``lab apparatus". 

It would be interesting if we could find a more natural interpretation of the quantity in Equation \ref{ShockwaveSize} as some sort of backreacted energy, or see if we can use spacelike left-right geodesics (which have well-defined endpoints rather than ending in the singularity) instead of null geodesics, but we leave this for future work\footnote{The eikonal phase we used is a measure of relative rapidity between the null geodesic traversed by the pulse, and the infalling particle. A similar definition should be possible for spacelike geodesics, but we used a null geodesic to obtain a clear connection to particle scattering. In the limit of large boosts, we expect the two definitions to agree; geodesic operators in AdS$_3$ have been shown to be proportional to null momenta in the light-cone limit \cite{ShockwaveOPE}. In the frame of the geodesic this can be interpreted as a large-boost limit for the matter sources that deform the geometry}.
\begin{figure}[!tbp]
  \centering
  \subfloat[Growth of a localized particle.]{\includegraphics[width=0.4\textwidth]{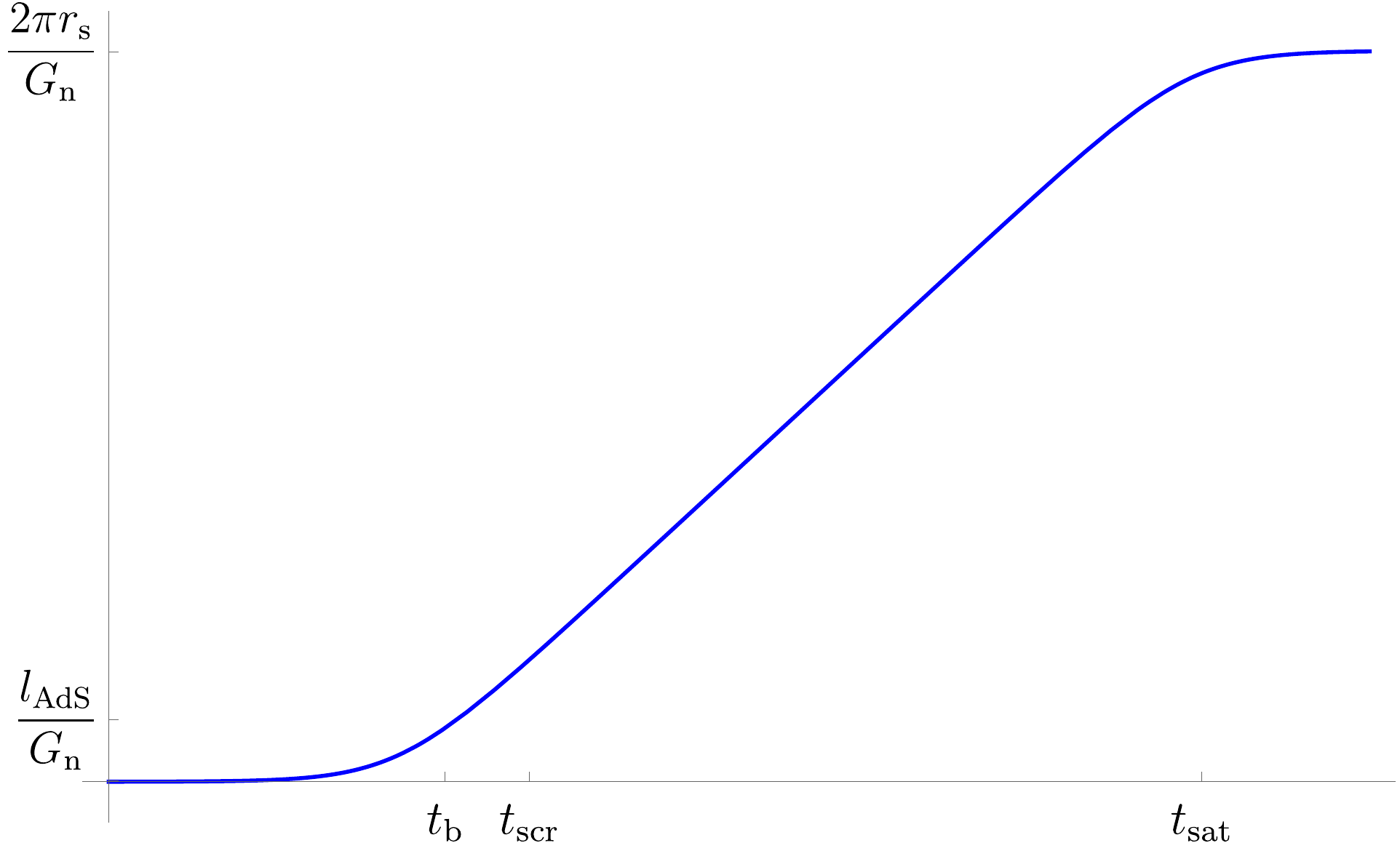}\label{fig:localized}}
  \hfill
  \subfloat[Growth of a spherical shell.]{\includegraphics[width=0.4\textwidth]{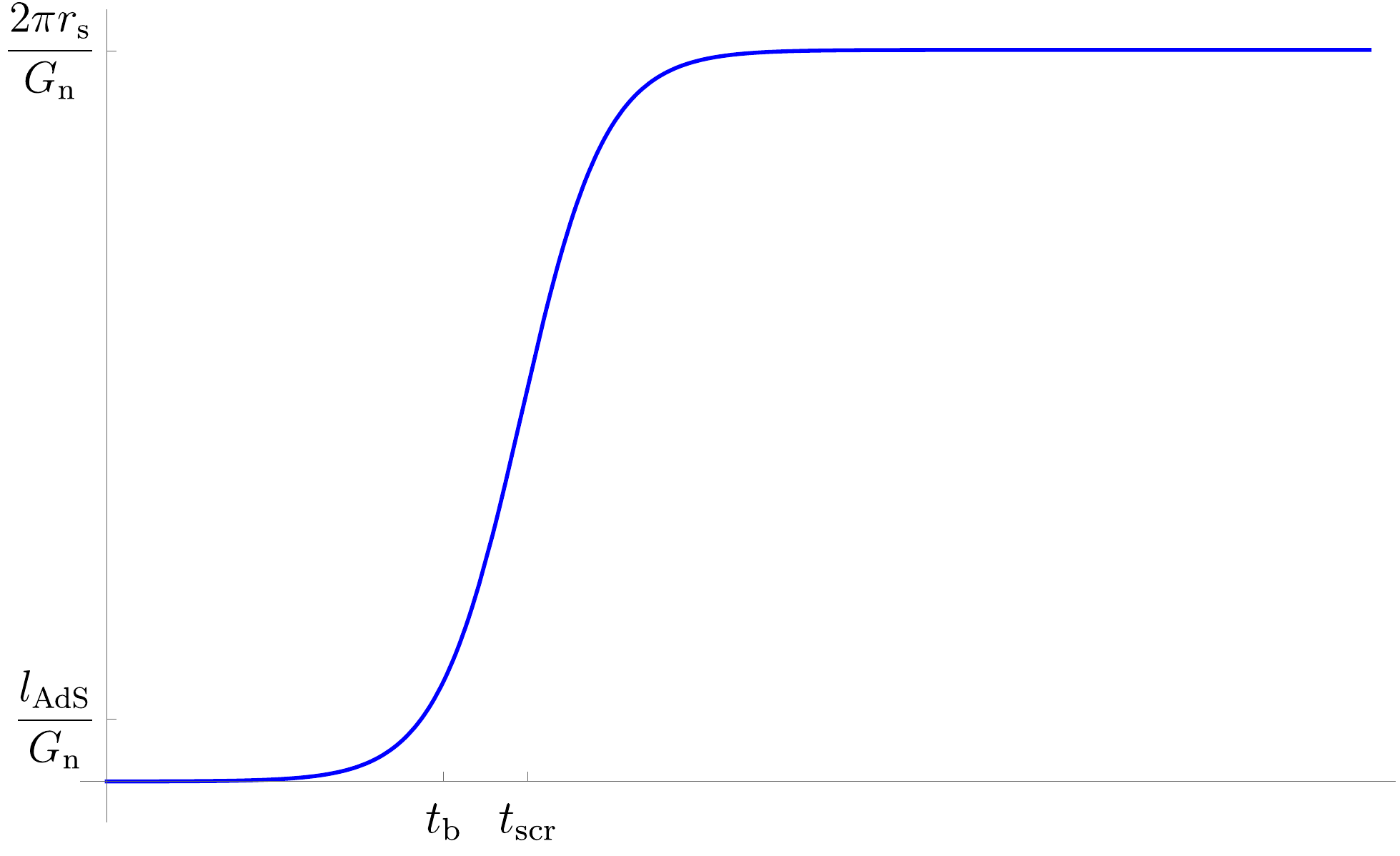}\label{fig:spherical}}
  \caption{Comparison of the size growth for a localized excitation and a spherical shell with equal energy. We have chosen the parameters to make it easier to visualize the exponential-linear growth transition.}
\end{figure}

Let's now find the growth patterns for our proposed operator size for localized excitations and spherical shells. If we consider a localized infalling excitation with Kruskal momentum $P_v$ at $\chi = 0$, then the operator size becomes 
\begin{equation}
\int \frac{d\chi}{l_{AdS}} \frac{P_v e^{-|\chi|/l_{AdS}}}{1 + \frac{G_N}{l_{AdS}}P_v e^{-|\chi|/l_{AdS}}}
\end{equation}
We plot the growth of this quantity in Figure \ref{fig:localized}. There is a clear change from an initial exponential growth to a linear growth, which occurs when $\frac{G_N}{l_{AdS}}P_v \sim 1$, and the size will be of order $l_{AdS}/G_N$. In AdS$_3$-Rindler, this linear growth will continue forever; as we increase $P_v \sim e^{t/l_{AdS}}$ the growth will come from values of $\chi \sim t$. However, in a BTZ black hole the values of $\chi$ are bounded by $2\pi r_s$ and the size levels off when $\frac{G_N}{l_{AdS}} P_v e^{-2\pi r_s/l_{AdS}} \sim 1$. It is easy to see that as $P_v \rightarrow \infty$ the operator size becomes
\begin{equation}
\int \frac{d\chi}{G_N} = \frac{2\pi r_s}{G_N} = S_{BH}
\end{equation}
Now let's repeat the same computation for a spherically symmetric shockwave with Kruskal momentum $P_v$. Then the operator size is
\begin{equation}
\int \frac{d\chi}{r_s} \frac{P_v}{1 + \frac{G_N}{r_s} P_v} = \frac{2\pi P_v}{1 + \frac{G_N}{r_s} P_v}
\end{equation}
The growth is plotted in Figure \ref{fig:spherical}, and it is immediately clear that it interpolates between an exponential growth at $P_v \ll \frac{r_s}{G_N}$ and a saturated value
\begin{equation}
\int \frac{d\chi}{r_s}\frac{r_s}{G_N} =  \frac{2\pi r_s}{G_N} = S_{BH}
\end{equation}
While it's good that the growth pattern matches our expectations, why should our proposed quantity be a natural candidate for an operator size? One reason is that the eikonal phase $\delta(s, \chi)$ measures the number of exchanged bulk gravitons between two particles that scatter with center-of-mass energy $s$ at impact parameter $\chi$. Indeed, at $\delta(s, \chi)\gg 1$ the eikonal series has a saddle \cite{ACV} which is dominated by diagrams of $\delta(s,\chi)$ loop order\footnote{In summary, the eikonal amplitude is obtained by resumming (crossed) ladder diagrams, and it can be written in impact parameter space as $\mathcal{A}_{eik}(s,\chi) \propto e^{i\delta(s,\chi)} - 1$. The exponential series is dominated by terms of order $\delta(s,\chi)$, and the interaction can be interpreted as a repeated exchange of $\delta(s,\chi)$ gravitons. This interpretation is supported by calculations of the momentum transfer, which are compatible with the exchange of $\delta(s,\chi)$ gravitons each carrying a small momentum of order $1/l_{AdS}$.}. From the boundary perspective, the exchanged bulk gravitons correspond to reparametrization modes that are responsible for OTOC growth\cite{ScramblingPhases}\cite{AdS3Reparametrizations}\cite{CFTQuantumChaos}. Eikonal phases of various scattering events are thus a natural probe of operator size, but it is important to dress the scattering events to the boundary so that they can be defined in a way that respects backreaction. 

Without carefully defining our pulses with respect to the boundary, their definition is ambiguous when the bulk geometry is not fixed. This is the same issue that arises in the bulk reconstruction of any field $\phi(X)$; the spacetime point $X$ must be well-defined even if the bulk geometry changes \cite{InsideOut}. The boundary remains invariant regardless of bulk perturbations, so it provides a natural reference point for the definition of any bulk observables. In our construction, we dressed our scattering events to the boundary by using pulses that are defined so that they will reach the boundary at times $t \leq 0$ regardless of the infalling shockwave. 

Among pulses that are dressed to the boundary, one can choose to define them in various ways. For example, one could consider pulses with fixed $P_v, E_0$ that are thrown in from the boundary regardless of the bulk geometry. Such pulses however will not lead to any saturating quantity. The pulses we chose have fixed energy $E_0$ and they always remain to the past of the $u = 0$ horizon, this is sufficient to place an upper bound on the eikonal phase

\begin{equation}\label{EikonalInequality}
\delta(s, \chi) < \beta E_0
\end{equation}
This inequality follows from the fact that the time-delay is
\begin{equation}
\Delta u = \frac{\partial \delta(s, \chi)}{\partial P_{u,pulse}} \sim \frac{\delta(s,\chi)}{P_{u,pulse}}
\end{equation}
and it must be smaller than $u_{pulse}$ for the pulse to avoid being pushed behind the horizon. By using the relation $P_u \sim \beta E/u$ that relates Kruskal momenta to the asymptotic energy, we obtain Inequality \ref{EikonalInequality}. This saturation suggests that this class of scattering events is suitable for defining an operator size for infalling shockwaves.

In all generality, we could consider any spherically symmetric combination of such scattering events, but in the shockwave limit we only need to consider radially moving pulses. Still, there is an ambiguity as to whether we should consider localized or delocalized pulses (e.g. $\chi$-momentum eigenstates). The eikonal phase measures the number of exchanged bulk gravitons only in impact parameter space (it's in impact parameter space that the eikonal series is dominated by a fixed loop order), which suggests the use of localized pulses. The above considerations uniquely lead to Equation \ref{ShockwaveSize} (plus the $u \rightarrow v$ term) if we take them seriously.

Of course, at the end of the day this is all just guesswork, but the resulting formula is very well behaved when it comes to shockwave geometries: it demonstrates both saturation and the expected growth patterns in a local theory. Additionally, since the eikonal phase measures the number of exchanged bulk gravitons, our operator size formula has an interesting bulk interpretation as measuring the average ``number of gravitons" that make up the gravitational field of an infalling particle (inasmuch as such a quantity can be defined). Of course, these aren't physical (on-shell) gravitons, but virtual gravitons that only make sense when they're measured by a probe. In vague terms (which we hope to make more precise in future work), this may be a measurement of the ``hydrodynamic cloud" \cite{HydroCloud} associated with an excitation.

\subsection{Universality of Operator Size, Entanglement and Complexity}\label{ComplexitySection}

As a final note, we want to go back to our results in Section \ref{AdSSection} and discuss the connection between different measures of growth in the boundary: operator size, entanglement and complexity. At leading order in $1/c$, we found that operator size is proportional to the growth of an ``average subsystem entropy", which due to a kinematic space formula is proportional to the complexity (volume) increase. To summarize, we write
\begin{equation}\label{SizeComplexityEntanglement}
\hat{S}_{|\Omega\rangle} \sim \frac{\delta V}{l_{AdS} G_N} \sim \int dx dy \frac{\partial^2 \mathcal{L}(x,0; y, 0)}{\partial x \partial y}  \frac{\delta\mathcal{S}(x,0 ; y,0)}{l_{AdS} G_N}
\end{equation}
where $\delta \mathcal{S} = \delta \mathcal{L}/G_N$ is the variation of entanglement entropy for the $(x,y)$ integral that lies on the static slice. To try to understand this equality, let us consider a cartoon version of the boundary CFT, which we represent as a discrete system with $N \gg 1$ degrees of freedom per site, and on which we can act with local 2-site operators. We take a basis of unitary gates $U_i$ which we use to define complexity\footnote{An assumption we will make is that the complexity of a state $U_1...U_k|\Omega\rangle$ is $C(|\Omega\rangle) + k$. In the large $N$ limit, we expect a typical gate $U_i$ to increase the complexity by 1 as long as we aren't in a state of near-maximal complexity.}, and a basis of fundamental operators $\mathcal{O}_i$ which we use to define operator size. We will take both sets to be made of 2-site operators that act on neighboring sites; 1-site gates (e.g. a phase rotation on a single site) can be built by taking a product of two 2-site gates that act on the same pair of sites $U_{i,i+1}V_{i,i+1}$. We will also assume a large number of gates (say, order $N$ as is the case in SYK), which we will write as $\mathcal{O}^a_{i,i+1}$ for a ``flavor" index $a$ and an operator acting on the sites $i,i+1$. We will assume that large-$N$ factorization holds for the fundamental operators.

First, we note that since a subregion $A$ of the boundary isn't maximally entangled with its complement, a typical 2-site gate that acts on both $A$ and its complements will increase the entanglement entropy $\mathcal{S}_A$. If we act with $k \ll N$ gates, then due to the large-$N$ limit we can apply a statistical reasoning and estimate that the entanglement entropy will typically increase by an $O(k)$ amount (up to a universal proportionality constant that shouldn't depend on the choice of gates). More generally, these gates will increase the average subsystem entropy defined by the rightmost side of Equation \ref{SizeComplexityEntanglement} by an $O(k)$ amount. Thus, if we want to compute the complexity of an operator $\mathcal{O}|\Omega\rangle = U_1U_2...U_k|\Omega\rangle$, we can instead compute the increase in the average subsystem entropy. This relation will fail if we consider an $O(N)$ number of gates, but as long as $k \ll N$ it should be valid. This explains the proportionality between the second and third terms in Equation \ref{SizeComplexityEntanglement}.

A somewhat more complicated argument can relate operator size to the average subsystem entropy. Suppose that we expand the action of a ``flavor typical" operator\footnote{We need some notion of non-correlation between the different terms in the expansion of $\mathcal{O}$ so that we can apply probabilistic arguments. Recall that in this toy model, our basis of fundamental operators has $O(N)$ elements per site.} on the vacuum in terms of monomials built from the fundamental operators
\begin{equation}
\mathcal{O}|\Omega\rangle = \sum_I c_I \Gamma_I|\Omega\rangle
\end{equation}
where $\Gamma_I$ is a monomial of the form
\begin{equation}
\prod_i \mathcal{O}^{a_i}_{k_i}
\end{equation}
Here, $a_i$ is a flavor index, and $k_i$ is a momentum index so that
\begin{equation}
\mathcal{O}^{a}_k = \sum_x \mathcal{O}^{a}_{x,x+1} e^{-ikx}
\end{equation}
The reason we chose to go in momentum space is to diagonalize 2-point functions. Furthermore, the size operator $\hat{S}_{|\Omega\rangle}$ cannot couple operators with different momenta
\begin{equation}
\langle (\mathcal{O}^a_k)^\dagger \hat{S}_{|\Omega\rangle} \mathcal{O}^{a'}_{k'} \rangle \propto \delta_{a,a'} \delta_{k,-k'}
\end{equation}
and a similar equality holds for the average subsystem entropy increase $\delta \mathcal{S}$\footnote{Of course, the entropy increase isn't generally given by an operator, but as evidenced by the RT formula it acts as an operator in states near the vacuum. We will assume that short monomials of the fundamental operators are compatible with the description of the entropy as an operator.}. However, in principle we can still have non-vanishing terms
\begin{equation}
\langle (\mathcal{O}^a_k)^\dagger (\mathcal{O}^b_l)^\dagger\hat{S}_{|\Omega\rangle} \mathcal{O}^{b'}_{l'} \mathcal{O}^{a'}_{k'} \rangle \propto \delta_{a,a'}\delta_{b,b'}\delta_{k+l, -k'-l'}
\end{equation}
These terms involve flavor repetitions in the monomials, and they are combinatorially disfavored in the large-$N$ limit so we can drop them. Thus, at leading order in $1/N$ only diagonal terms contribute to the expectation values of $\hat{S}_{|\Omega\rangle}, \delta \mathcal{S}$ and we write
\begin{equation}\label{EntanglementSize}
S_{|\Omega\rangle}(\mathcal{O}) = \frac{\sum_I |c_I|^2 S_{|\Omega\rangle}(\Gamma_I)}{\sum_I |c_I|^2}, \quad \delta \mathcal{S}(\mathcal{O}) = \frac{\sum_I |c_I|^2  |\delta\mathcal{S}(\Gamma_I)}{\sum_I |c_I|^2}
\end{equation}
In order to establish proportionality between the two quantities, it suffices to do so term by term.

As in SYK, due to large-$N$ factorization the size of a state
\begin{equation}
\prod_i \mathcal{O}^{a_i}_{k_i}|\Omega\rangle
\end{equation}
is the ``naive size", i.e. the number of $\mathcal{O}$'s that appear in the product. Since each $\mathcal{O}^{a_i}_{k_i}$ is a linear combination of 2-qubit operators, we expect it to increase the average subsystem entropy by an $O(1)$ amount (just as we argued previously for complexity, this arises because the various subregions are far from maximally entangled). Since we have no flavor repetition, the total increase in the entropy is obtained by summing over the increase due to each $\mathcal{O}^a_k$. Each such increase is flavor-independent, but it can be momentum-dependent. If we write the size of $\mathcal{O}^a_k$ as $f(k)$, the size of each monomial is
\begin{equation}
\delta \mathcal{S}(\prod_i \mathcal{O}^{a_i}_{k_i}) = \sum_i f(k_i)
\end{equation}
This relation between size and entropy isn't universal, but in the limit when we consider large monomials (but still small compared to $N$), we can apply the central limit theorem and write
\begin{equation}
\delta \mathcal{S}(\prod_{i=1}^n \mathcal{O}^{a_i}_{k_i}) = n \bar{f}
\end{equation}
The size $n$ we need for this to be a good approximation may be dependent on the ratio of the system's spatial size to the lattice size. For example, if the values of $f(k)$ increase with momentum, then the variance will increase as we decrease the lattice size and the central limit will converge more slowly. A better bound may be possible but for now this will suffice. Up to an overall numerical coefficient $\bar{f}$ which may depend on the system/lattice size (but not on $N$), we have established
\begin{equation}
\delta \mathcal{S}(\Gamma_I) \sim S_{|\Omega\rangle}(\Gamma_I)
\end{equation}
holds for all large (but not order $N$) monomials, and Equation \ref{EntanglementSize} ensures that $S_{|\Omega\rangle}(\mathcal{O})\sim \delta \mathcal{S}(\mathcal{O})$.

Perturbatively, we see that there is an expected universality in these measures of growth in the boundary theory. Ultimately, what all these measures do is count the number of gates (as long as there's not too many of them). In the AdS$_3$ vacuum, the first law of entanglement applied to the average subsystem entropy relates both complexity and operator size to the Hamiltonian.

Of course, complexity is a more useful measure not in the perturbative regimes that we are considering here, but in late time regimes where other measures of complexification (e.g. the entanglement entropy growth) have already saturated. When we start using a large number of gates/operators, the above large-$N$ arguments fail, and these three quantities start to differ. The most stark difference is between complexity and operator size. Their definitions are almost identical, except that complexity only allows products of gates (and requires approximate, not exact equality), while operator size allows for both sums and products of gates. When it comes to measuring the complexity/size of small perturbations to the vacuum, it seems that the banning of linear combinations only costs an $O(1)$ proportionality factor. But as we start looking at states far from the vacuum, this restriction makes it increasingly difficult to reproduce these states, as evidenced by the fact that complexity can grow to $O(e^N)$ values while operator size only grows to $O(N)$ values.

\section{Discussion}

We have formulated a state-dependent definition of operator size and related it to a positive semi-define operator $\hat{S}_{|\Psi\rangle}$ which we have explicitly computed for a large class of SYK states. By postulating that in holographic theories $\hat{S}_{|\Psi\rangle}$ must be built by operators that appear in the growing part of the OPE, we have deduced from symmetry arguments that the vacuum size $\hat{S}_{|\Omega\rangle}$ must be proportional to the Hamiltonian $H_{CFT}$. This allowed us to derive an expression for the operator size in AdS-Rindler and its quotients, the BTZ black holes. This expression captured only the leading behavior in the $1/c$ expansion. We conjectured that higher $1/c$ corrections to operator size are captured by a spatial average of the eikonal phases associated to a class of scattering events that are carefully defined with respect to the boundary (so that they'll be well-defined despite backreaction). 

These conjectured corrections predict an interesting pattern for operator growth past the exponential region: localized excitations transition to a linear growth period before saturating, while spherically symmetric shells have keep growing exponentially until they saturate (see Figures \ref{fig:localized}, \ref{fig:spherical}). This behavior is in line with general expectations from operator growth in a local Hamiltonian with $N \gg 1$ degrees of freedom per site (where we assume a $k$-local all-to-all interaction in each site). This computation should be possible to do explicitly for higher-dimensional versions of the SYK model, where the properties of Majorana fermions still allow an explicit writing of the size operator\footnote{For example, if we include a flavor-diagonal hopping term in a chain of SYK models, then the operator size is identical to Equation \ref{ThermalSize2} except that we replace the SYK Majorana fermions with spatial momentum modes $\psi^{j,k}$.}

Below, we discuss a collection of issues and future directions that could be of interest.

\subsection*{Operator Size and Error Correction}

We found that the operator size in vacuum AdS is simply given by the Hamiltonian, and thus the size of a particle moving through the bulk will be constant regardless of its location. This appears to be in some tension with the picture of AdS/CFT as an error correcting code \cite{ErrorCorrection}, where operators deep in the IR are better protected from boundary erasures, and one would assume they need to be larger operators on the boundary.

Recall that the size of an operator $\mathcal{O}$ is given by a minimization problem, where we choose the ``shortest" representation of $\mathcal{O}|\Psi\rangle$. Still, it can have a large number of representations; there's no obvious relation between the size of an operator and the number of representations, and neither is there a relation between the operator size and spatial size (before finite $N$ effects come into play).

Suppose we have a bulk operator $\phi(X)$ and we wish to reconstruct it in a subregion $A$. As long as $X$ is in the Rindler wedge of $A$, we can reconstruct $\phi(X)$ on the boundary $C(A)$ of the Rindler wedge\footnote{See \cite{BulkSYK} for a treatment of the operator size of bulk fields in the SYK model.}
\begin{equation}
\phi(X) = \int_{C(A)} dx dt \, K(X;x, t) \mathcal{O}(x, t)
\end{equation}
An important point is that the operators $\mathcal{O}(x,t)$ are evolved with the Rindler Hamiltonian, and thus they are highly energetic with regards to the global Hamiltonian. If $X$ is near the horizon of the Rindler wedge, this appears to suggest that $\phi(X)$ will be written in terms of large operators, and thus its size will be large as well. However, this expectation is naive; the size of any representation of $\phi(X)$ will be given by the global energy of $\phi(X)|\Omega_{bulk}\rangle$. A sum of large operators need not necessarily be a large operator, just as a sum of high-energy modes can have low energy.

There is one sense in which the position of $\phi(X)$ and its size are correlated. For a finite-energy field to be localized in the near-horizon region of the Rindler wedge, it must have a sufficiently short wavelength and thus a sufficiently high energy. More concretely, in AdS-Rindler coordinates, localizing a wavepacket in the region $0 < u < u_0$ requires a Kruskal momentum that's $P_u \gtrsim 1/u_0$. Thus, it isn't possible to have a small operator ``deep in the bulk". The closer we want to localize a bulk operator near the horizon, the higher energy we'll need, and the higher it's size. 

Similarly, in global coordinates one needs a high energy to localize a smeared field near the center of AdS space. So while there isn't a direct relation between position and size, it is clear that any field that enjoys a large degree of protection from boundary erasures (i.e. is localized near the center of AdS space) must have a large operator size. From a boundary perspective, we can summarize this as follows: a large operator need not be well-protected from localized erasures, but a well-protected operator must be large.

\subsection*{Operator Size and Backreaction}

The eikonal-corrected operator size proposed in Equation \ref{ShockwaveSize} has the convenient property that it automatically stays within the eikonal regime regardless of the infalling particle. This allows it to be defined well past the scrambling time, in analogy with the $SL(2)$ charges of \cite{NearHorizon}. As we have repeatedly mentioned, Equation \ref{ShockwaveSize} is little more than an educated guess, and we would like to put its formulation on more solid ground. The interpretation of Equation \ref{ShockwaveSize} as counting the ``number of gravitons" that make up the gravitational field is appealing, and we hope to relate this to the hydrodynamic/reparametrization treatments of \cite{HydroCloud,AdS3Reparametrizations,ChaosEFT, CFTQuantumChaos}. 

One more direction we would like to understand is how to modify Equation \ref{ShockwaveSize} to properly count the size of shockwaves whose mass is a small (but finite) fraction of the black hole mass. The scattering events stay within the eikonal regime and in the near-horizon region as long as the shockwave mass is $M= \epsilon l_{AdS}/G_N$ for some $\epsilon \ll 1$ (but parametrically $O(1)$ in the $1/N$ expansion). Equation \ref{ShockwaveSize} thus remains under perturbative control, but it computes a maximum size $S_{BH}$ regardless of $M$. Given our intuition from scrambling, we would expect that the saturated size will be equal to the perturbed black hole entropy $S_{BH} + \delta S_{BH}$. One ``phenomenological" way to achieve that is to define the change the measure $d\chi/G_N$ of the integral (which captures the density of pulses per transverse area) so that it respects backreaction. In the presence of a massive shockwave, the BTZ geometry in the future of the shockwave is perturbed and it can be described by a new set of Kruskal coordinates $(u', v', \chi')$ which give the metric the same form as 
Equation \ref{AdSRindlerMetric} with an identification $\chi' \sim \chi' + 2\pi r_s'$ (with $r_s'$ being the perturbed black hole radius). A surface $u'v' = const$ can by sending in geodesics from the boundary with fixed renormalized length, so it can be defined in a gauge invariant manner. In order to formulate a well-defined measurement of the eikonal phase, we must specify the asymptotic behavior of the pulses as $u'v' \rightarrow -1$. We could choose to have a fixed number (or measure, in a continuum limit) of pulses to cross each such surface, or we can instead specify the density of pulses that cross a unit area on the surface $u'v' = const$. If we choose the latter, then we define the spatial density of the pulses that cross $u'v' = const$ to scale as 
\begin{equation}
\frac{1+u'v'}{1-u'v'}\frac{1}{G_N}
\end{equation}
This is inversely proportional to the area of the surface and it gives a finite total measure for the pulses. However, the measure will be dependent on the bulk geometry, and it will give a factor $d\chi'/G_N$ in the eikonal phase integral. The average eikonal phase under this measure is thus
\begin{equation}
\int \frac{d\chi'}{G_N} \frac{\Delta u(\chi')}{1+\Delta u(\chi')}
\end{equation}
This formula is identical to Equation \ref{ShockwaveSize}, except that it is defined with respect to the backreacted coordinates, and due to the increased periodicity of $\chi'$ it will saturate at a value $S_{BH} + \delta S_{BH}$. It is interesting that there isn't a need for any interaction beyond the eikonal to capture these corrections; they instead come the modification of the asymptotic boundary conditions of our measurement process due to the increased mass of the black hole (in other words, the holographic dictionary changes, and thus we get a different bulk interpretation for the same boundary operator).

One could ask, why don't we instead multiply Equation \ref{ShockwaveSize} by an area operator (defined on states near $|\beta\rangle$) to capture this increase? One reason is just that we wanted a ``natural" definition in terms of an appropriate measurement of the gravitational field of the infaller, which we hope will map to an intuitive boundary interpretation. Another reason is that in order to assign the horizon an area operator in a time-dependent geometry, one needs to dress the horizon to the boundary\footnote{The black hole horizon area will no longer be given by the entanglement entropy of the right boundary once we add a perturbation. Entanglement entropies are easy to formulate as operators, but that's not the case for the area of arbitrary surfaces.} to define it in a gauge-invariant way. This is no more straightforward than our construction which relied on dressing surfaces in the UV to the boundary.

\subsection*{Rotating, Hyperbolic and Time-Shifted Black Holes}

Our construction for $\hat{S}_{|\Omega\rangle}$ equally applies well to higher dimensions, and so does the generalization to AdS-Rindler. While in 3 dimensions one can construct the most general class of static black holes (BTZ black holes) from quotients of AdS-Rindler, in higher dimensions one can only do so for hyperbolic black holes that are quotients of the spacetime
\begin{equation}
ds^2 = l_{AdS}^2\frac{-4 du dv}{(1+uv)^2}+\frac{(1-uv)^2}{(1+uv)^2}dH^2_{d-1}
\end{equation}
by a discrete isometry group of the hyperbolic space $H_{d-1}$. Investigating operator size for these geometries should be a straightforward extension of the 3-dimensional case (see \cite{HyperbolicScrambling} for chaos calculations of OTOCs and the butterfly velocity in these geometries). Another interesting case would be to try to generalize our construction to rotating BTZ geometries.

The AdS$_3$ vacuum can be re-written in ``rotating AdS-Rindler" \cite{RotatingAdSRindler} coordinates, from which one can obtain the rotating BTZ black hole by taking a quotient on the angular coordinate \cite{RindlerAdSCFT}. This suggests we could investigate operator growth in for rotating BTZ black holes. These geometries have a different effective temperature $\beta_+, \beta_-$ for the left/right moving modes, which gives some interesting effects on OTOC growth \cite{BTZDynamics}\cite{ChaosBoundRotating}. 

A further generalization we would like to understand is operator size on time-shifted TFDs. The results we obtained in SYK (see Equation \ref{TimeShiftedSize}) appear like they should generalize to higher dimensions. In line with the perturbative agreement of operator size and complexity, we expect that the size of a shockwave for a time-shifted BTZ should be
\begin{equation}
S_{|\beta(t_R, t_R)\rangle}(\mathcal{O}_R(-t)) \sim \frac{\Delta_{\mathcal{O}}}{\sin(\frac{2\pi\tau}{\beta})} \frac{\cosh(\frac{2\pi}{\beta}t)}{\cosh(\frac{2\pi}{\beta}t_R)}
\end{equation}
for $t_R \lesssim t_{scr}$, with a transition occuring as $t_R \gtrsim t_{scr}$. By taking a quotient that maps the left side to the right side, this would allow us to understand operator in single-sided boundary state black holes \cite{EternalBlackHoles, EscapingInteriors}. Currently, we can construct the operator size of these states for $t_R = 0$, but a more general time-dependence (in analogy with Equation \ref{PureStateSize}) would be desirable. As in SYK, we expect that typical black hole microstates will have an operator size that is ``identical" to the thermofield double $|\beta\rangle$ and is thus given by $P_u + P_v$ in the near-horizon region.

\subsection*{Size Saturation and Microscopics}

It is still not clear however why size saturation should ever happen in a field theory. While in fermionic theories saturation is obvious due to the non-repetition of flavors, there is no such restriction in a CFT. Maybe a clearer relation of operator size to emergent hydrodynamics would help, or perhaps one could hope to gain a better ``microscopic" description of operator size. 

A direct approach based on explicitly constructing the operator $\hat{S}_{|\Psi\rangle}$ from the local fields of a CFT will probably be difficult. Besides the usual problems for an interacting theory (e.g. the need for a temporal smearing of local fields to yield well-defined operators makes the choice of fundamental operators ambiguous), there may be additional difficulties with creating a gauge-invariant operator size.

It is also possible that for interacting theories, there is no unique definition of an operator size, but there's always an explicit dependence on some cutoff or smearing procedure. If that's the case, then we'd (optimistically) expect that our current results capture the universal growth behavior. To give an RG analogy, the definition of operator size may have some UV (small operator) artifacts, but it may still give a universal IR (large operator) behavior. By small and large operators, we don't mean compared to $N$, but compared to a cutoff-dependent quantity.

From a holographic perspective, we can imagine that the bulk dual of operator size has an explicit dependence on the cutoff surface. In SYK, the cutoff surface is physical and unambiguous, but in CFTs it is an arbitrary regulator. In analogy with the renormalization goroup, we can hope that even if the behavior of operator size near the boundary (UV) is cutoff-dependent, when we look at excitations deep in the bulk (IR) the cutoff-dependence fades away and we obtain a universal behavior.

\section*{Acknowledgments}

We wish to thank Bartlomiej Czech, Javier Magan, Yuri Lensky, Xiaoliang Qi, Phil Saad, Stephen Shenker, Milind Shyani, Alexandre Streicher, Leonard Susskind, Zhenbin Yang and Ying Zhao for helpful comments and discussions. We especially thank Douglas Stanford for extensive discussions.

\newpage

\appendix

\section{Operator Size for Bosonic Systems}\label{Bosons}

Operator size is straightforward to define for fermionic systems because every operator can be uniquely decomposed into monomials of anticommuting Majorana fermions. For a finite-dimensional Hilbert space of dimension $2^N$, it is clear that we need $2N$ Majorana fermions/Pauli matrices to generate the algebra of operators, and a choice of such fundamental operators uniquely determines our definition of operator size. But what happens if we consider the simplest bosonic system, the harmonic oscillator?

Here we only have a single flavor (unlike fermionic systems where we have $N$ distinct Majorana fermions), but an infinite number states. We could have taken the trivial definition of operator size to say that any operator acting on a single harmonic oscillator has size 1 since it only affects one ``site", but such a definition wouldn't be particularly useful. Ideally, we would want there to be a notion of how difficult it is to create a given operator starting from a set of fundamental operators. For a harmonic oscillator, a natural choice of such operators would be $a, a^\dagger$.

By expanding in terms of monomials in $a, a^\dagger$, the definition of operator size then seems like it could proceed analogously to \ref{AverageSize}, but there are two important complication. The first issue is that it is difficult to define the normalization of an operator. The normalization for $a$ should be given by

\begin{equation}
\text{tr}(a^\dagger a)
\end{equation}
which diverges. We could try to put a cutoff on the oscillator states, in which case
\begin{equation}
\text{tr}(a^\dagger a) \simeq \frac{N_{UV}^3}{3}
\end{equation}
But then the problem is that the normalization of $a^\dagger a$ is
\begin{equation}
\text{tr}(a^\dagger a a^\dagger a) \simeq \frac{N_{UV}^5}{5}
\end{equation}
which has a higher power of the cutoff. If we wished to compute the size of $a^\dagger a + \epsilon (a^\dagger a)^k$, then there would be the issue that regardless of how small $\epsilon$ is, the size of this operator would be $k$ as we take $N_{UV} \rightarrow \infty$. This problem will be particularly pronounced if we try to compute the growth of an operator, say $a(t)$, when evolving with a Hamiltonian $a^\dagger a + \lambda a^\dagger a^\dagger a a$. For any finite time evolution, $a(t)$ will be an infinite series including very high powers of $a, a^\dagger$. Terms of $k$-th degree will be suppressed by small coefficients of order
\begin{equation}
\frac{(\lambda t)^k}{k!}
\end{equation}
but terms with high $k$ will nevertheless dominate in the limit $N_{UV} \rightarrow \infty$ and $a(t)$ will have infinite size regardless of how small $t$ is.

The presence of a UV cutoff is crucial to obtain a well-defined operator size, and the resulting operator size sensitively depends on the cutoff. Thus, it is impossible to define a truly ``state-independent" operator size, and we can only define operator size for an ensemble of states (in this case, the states under the cutoff). Thus, we will only attempt to define a state-dependent size. 

For any state $|\Psi\rangle$ whose levels aren't arbitrarily high, the normalization of any operator $O$ will be finite and equal to $\langle \Psi |O^\dagger O|\Psi\rangle$. However, we should note that we can take normalized states that have divergent values even for ``simple" operators. For example,
\begin{equation}
|\Psi\rangle = \sum_{n=1}^\infty \frac{1}{n}|n\rangle
\end{equation}
has a finite normalization $\langle \Psi|\Psi\rangle = \sum \frac{1}{n^2} = \frac{\pi^2}{6}$ but it also has a divergent expectation value for $a^\dagger a$. This is a well-known issue of distributions that can have a finite normalization, but divergent moments above a certain power. However, as long as we ensure that the level-distribution of $|\Psi\rangle$ is bounded by
\begin{equation}\label{Reasonable}
|c_n|^2 \lesssim C_1 e^{-\beta n}
\end{equation}
for some $\beta > 0$ and some constant $C_1$, then we can guarantee that all monomials of $a, a^\dagger$ have a finite expectation value that is at most $C_2 n_0^k$ for some positive $n_0 >0$ and $k$ being the degree of the monomial. This also ensures that the time-evolution of operators is well-behaved, in the sense that high-order monomials appearing in the expansion of $a(t)$ will be highly-suppressed (even after we account for the normalization), and we expect the ``naive operator size" given by Equation \ref{UnnormalizedSize} to be finite and continuous as a function of time (e.g. we won't see pathological behavior where the size of $a(t)$ jumps to a large value the moment $t$ becomes non-zero).

Overall, we see that ``reasonable" states are expected to give a well-behaved operator size if we follow Equation \ref{AverageSize} with the appropriate normalizations, but there are two more issues to consider. First, is it clear that every operator can be written in terms of $a, a^\dagger$? While it's not obvious that this is true, what we can say is that the action of $\mathcal{O}|\Psi\rangle$ can be replicated as a convergent sum of monomials in $a, a^\dagger$ as long as $\mathcal{O}|\Psi\rangle$ is a reasonable state (in the sense of satisfying \ref{Reasonable}). The proof is a straightforward but cumbersome exercise in real analysis, but let us briefly note how this would work if $|\Psi\rangle$ and $\mathcal{O}|\Psi\rangle$ are both states whose highest level is finite. In this case, we can expand
\begin{equation}
\mathcal{O}|\Psi\rangle = \sum_{n=1}^{n_{max}} b_n |n\rangle
\end{equation}
which can be obtained from $|\Psi\rangle$ by acting with the operator

\begin{equation}
\sum_{i=1}^{n_{max}} b_n \frac{(a^\dagger)^n}{\sqrt{n!}} \frac{a^{n_{\Psi}}}{c_{n_\Psi} \sqrt{n_{\Psi}!}}
\end{equation}
where $n_{\Psi}$ is the highest level of $|\Psi\rangle$ and $c_{n_\Psi}$ is its coefficient. What the above operator does is taken $|\Psi\rangle$ down to the vacuum $|0\rangle$, and then it builds $\mathcal{O}|\Psi\rangle$ by acting on $|0\rangle$ with creation operators. From the above, it is clear how any reasonable operator (e.g. something that's not $e^{\beta H}$ for some $\beta >0$) can be given a finite size representation in terms of $a, a^\dagger$. 

However, $a$ and $a^\dagger$ do not commute, and thus there are many ways to represent the same operator $\mathcal{O}$ in terms of monomials. As a trivial example, the operator $a a^\dagger$ can be rewritten as $a^\dagger a + 1$. If we wanted to find the size of $a a^\dagger$ with respect to the vacuum state, then this representation would give size 2, while the representation $a^\dagger a + 1$ would give size 0 since $a^\dagger a$ annihilates the vacuum. So in order to have a proper definition of operator size, it is important that we minimize over all possible representations in terms of $a, a^\dagger$.

Since we know that a finite size representation exists, we know that an infimum over operator sizes does exist, but it's not clear that it can be achieved since there is a potentially infinite number of such representations. For reasonable states, one can show that there is a convergent sequence that reaches the infimum (in an $L_2$ norm), so the minimization can indeed be achieved. The main point is that due to the exponential falloff of the coefficients $|c_n|^2$ of $|\Psi\rangle$, one can effectively cut off large combinations of $a^\dagger, a$ in any sequence that approaches the infimum with only an exponentially small correction in size. By carefully bounding these corrections, a convergent sequence that attains the infimum can be attained. 

With that said, we can now follow a similar argument to Section \ref{Construction} to show that the operator size defines an inner product on reasonable states $\mathcal{O}|\Psi\rangle$, and thus we can write it in the form
\begin{equation}
S_{|\Psi\rangle}(\mathcal{O}) = \frac{\langle \Psi| \mathcal{O}^\dagger \hat{S}_{|\Psi\rangle} \mathcal{O}|\Psi\rangle}{\langle \Psi|\mathcal{O}^\dagger \mathcal{O}|\Psi\rangle}
\end{equation}
where $\hat{S}_{|\Psi\rangle}$ is a positive semi-definite, Hermitian operator that annihilates $|\Psi\rangle$ and no other state. This equation, entirely analogous to the fermionic one, allows us to formulate operator size for field theories. In practice however, this derivation was just to give us a sense of certainty that operator size can indeed be written as an observable; guessing the exact observable is the hard part.

\bibliographystyle{JHEP}

\begin{thebibliography}{1}


\bibitem{ButterflyEffect}
  S. H. Shenker, D. Stanford, ``Black holes and the butterfly effect," JHEP 1403 (2014) 067
  
\bibitem{EntanglementTsunami}
  H. Liu, S. J. Suh, ``Entanglement Tsunami: Universal Scaling in Holographic Thermalization," Phys.Rev.Lett. 112 (2014) 011601
  
\bibitem{StringyEffects}
  S.~H.~Shenker, D.~Stanford, ``Stringy effects in scrambling," JHEP 1505 (2015) 132
  
\bibitem{ComplexityShocks}
  D.~Stanford, L.~Susskind, ``Complexity and Shock Wave Geometries,"Phys.Rev. D90 (2014) no.12, 126007
  
\bibitem{ChaosBound}
  J. Maldacena, S. H. Shenker, D. Stanford, ``A bound on chaos," JHEP 1608 (2016) 106
  
  
\bibitem{ThingsFall}
  L.~Susskind, ``Why do Things Fall?," arxiv:1802.01198 [hep-th].
  
\bibitem{FallingCharged}
  A.~Brown, H.~Gharibyan, A.~Streicher, L.~Susskind, L.~Thorlacius, Y.~Zhao ``Falling Toward Charged Black Holes," Phys.Rev. D98 (2018) no.12, 126016
  
  
\bibitem{ComplexityLaws}
  L.~Susskind, ``Complexity and Newton's Laws," arXiv:1904.12819 [hep-th]  
  
\bibitem{Streicher-Qi}
  X.-L.~Qi, A.~Streicher, ``Quantum Epidemiology: Operator Growth, Thermal Effects, and SYK," JHEP 1908 (2019) 012
  
\bibitem{OperatorGrowth}
  D. A.~Roberts, D.~Stanford, A. Streicher, ``Operator growth in the SYK model," JHEP 1806 (2018) 122

\bibitem{LucasThermal}
  A.~Lucas, ``Operator Size at Finite Temperature and Planckian Bounds on Quantum Dynamics," Phys. Rev. Lett. 122, 216601

\bibitem{NearHorizon}
  H.~W.~Lin, J.~Maldacena, Y.~Zhao, ``Symmetries Near the Horizon," JHEP 1908 (2019) 049



\bibitem{GJW}
  P.~Gao, D.~L.~Jafferis, A.~C.~Wall, ``Traversable Wormholes via a Double Trace Deformation," JHEP 1712 (2017) 151

\bibitem{DivingTraversable}
  J.~Maldacena, D.~Stanford, Z.~Yang, ``Diving into traversable wormholes," Fortsch.Phys. 65 (2017) no.5, 1700034


\bibitem{EternalTraversable}
  J. Maldacena, X.-L. Qi, ``Eternal traversable wormhole," arXiv:1804.00491 [hep-th] 

\bibitem{4DTraversable}
  J.~Maldacena, A.~Milekhin, F.~Popov, ``Traversable wormholes in four dimensions," arXiv:1807.04726 [hep-th]

\bibitem{MaldacenaKourkoulou}
  I.~Kourkoulou, J.~Maldacena, ``Pure states in the SYK model and nearly-\textit{AdS}$_2$ gravity," arXiv:1707.02325 [hep-th]

\bibitem{EscapingInteriors}
  A.~Almheiri, A.~Mousatov, M.~Shyani, ``Escaping the Interiors of Pure Boundary-State Black Holes," arXiv:1803.04434 [hep-th]

\bibitem{InteriorTypicalMicrostate}
  J.~de Boer, R.~Van Breukelen, S.~F.~Lokhande, K. Papadodimas, E. Verlinde, ``On the interior geometry of a typical black hole microstate," JHEP 1905 (2019) 010

\bibitem{SelfSupporting}
  Z. Fu, B. Grado-White, D. Marolf, ``A perturbative perspective on self-supporting wormholes," Class.Quant.Grav. 36 (2019) no.4, 045006

\bibitem{RotatingTraversable}
  E.~Caceres, A. S. Misobuchi, M.-L. Xiao, ``Rotating traversable wormholes in AdS," JHEP 1812 (2018) 005

\bibitem{ShockwaveOPE}
  N.~Afkhami-Jeddi, T.~Hartman, S.~Kundu, A.~Tajdini, ``Shockwaves from the Operator Product Expansion," JHEP 1903 (2019) 201
  
\bibitem{EntropyEoM}
  B.~Czech, L.~Lamprou, S.~McCandlish, B.~Mosk, J.~Sully, ``Equivalent Equations of Motion for Gravity and Entropy," JHEP 1702 (2017) 004
  
\bibitem{BuildTFD}
  W.~Cottrell, B.~Freivogel, D.~M.~Hofman, S.~F.~Lokhande, ``How to Build the Thermofield Double State," JHEP 1902 (2019) 058

\bibitem{SoftModeSYK}
  A. Kitaev, S. J. Suh, ``The soft mode in the Sachdev-Ye-Kitaev model and its gravity dual," JHEP 1805 (2018) 183
 
  
\bibitem{MaldacenaStanfordYang}
  J.~Maldacena, D.~Stanford, Z.~Yang, ``Conformal symmetry and its breaking in two dimensional Nearly Anti-de-Sitter space," PTEP 2016 (2016) no.12, 12C104



\bibitem{RindlerAdSCFT}
  M.~Parikh, P.~Samantray, ``Rindler-AdS/CFT," JHEP 1810 (2018) 129

\bibitem{RindlerQuantumGravity}
  B.~Czech, J.~L.~Karczmarek, F.~Nogueira, M.~Van Raamsdonk, ``Rindler Quantum Gravity," Class.Quant.Grav. 29 (2012) 235025

  
\bibitem{ScramblingPhases}
  T.~Anous, J.~Sonner, ``Phases of scrambling in eigenstates," SciPost Phys. 7 (2019) 003


\bibitem{GravityDualBoundaryCausality}
  N.~Engelhardt, S.~Fischetti, ``The Gravity Dual of Boundary Causality," Class.Quant.Grav. 33 (2016) no.17, 175004

  
\bibitem{EternalBlackHoles}
  J.~Maldacena, ``Eternal black holes in Anti-de-Sitter," JHEP 0304 (2003) 021
  
\bibitem{RotatingAdSRindler}
  M.~Parikh, P.~Samantray, E.~Verlinde, ``Rotating Rindler-AdS Space," Phys.Rev. D86 (2012) 024005

\bibitem{KinematicComplexity}
  R.~Abt, J.~Erdmenger, M.~Gerbershagen, C.~M.~Melby-Thompson, C.~Northe, ``Holographic Subregion Complexity from Kinematic Space," JHEP 1901 (2019) 012

\bibitem{EntwinementEmergence}
  V.~Balasubramanian, B.~D.~Chowdhury, B.~Czech, J.~ de Boer, ``Entwinement and the emergence of spacetime," JHEP 1501 (2015) 048

\bibitem{ComplexityShockwaves}
  D.~Stanford, L.~Susskind, ``Complexity and Shock Wave Geometries," Phys.Rev. D90 (2014) no.12, 126007
  
\bibitem{ANEC}
  T.~Hartman, S.~Kundu, A.~Tajdini, ``Averaged Null Energy Condition from Causality," JHEP 1707 (2017) 066

\bibitem{ACV}
  D.~Amati, M.~Ciafaloni, G.~Veneziano, ``Classical and Quantum Gravity Effects from Planckian Energy Superstring Collisions," Int.J.Mod.Phys. A3 (1988) 1615-1661


\bibitem{GravityDualsModular}
  D.~L.~Jafferis, S.~J.~Suh, ``The Gravity Duals of Modular Hamiltonians," JHEP 1609 (2016) 068


\bibitem{RelativeEntropyBulk}
 D.~L.~Jafferis, A.~Lewkowycz, J.~Maldacena, S.~J.~Suh, ``Relative entropy equals bulk relative entropy," JHEP 1606 (2016) 004

\bibitem{NewYorkTime}
  A.~Belin, A.~Lewkowycz, G.~Sarosi, 	``Complexity and the bulk volume, a new York time story," JHEP 1903 (2019) 044

\bibitem{LocalizedShocks}
  D.~A.~Roberts, D.~Stanford, L.~Susskind, ``Localized shocks," JHEP 1503 (2015) 051
  
\bibitem{MirrorOperators}
  K. Papadodimas, S. Raju, ``An Infalling Observer in AdS/CFT," JHEP 1310 (2013) 212

\bibitem{HydroCloud}
  M.~Blake, H.~Lee, H.~Liu, ``A quantum hydrodynamical description for scrambling and many-body chaos," JHEP 1810 (2018) 127

\bibitem{AdS3Reparametrizations}
  J.~Cotler, K.~Jensen, ``A theory of reparameterizations for AdS$_3$ gravity," JHEP 1902 (2019) 079

\bibitem{ChaosEFT}
  F.~M.~Haehl, M.~Rozali, ``Effective Field Theory for Chaotic CFTs," JHEP 1810 (2018) 118
  
\bibitem{CFTQuantumChaos}
  G. Turiaci, H. Verlinde, ``On CFT and Quantum Chaos," JHEP 1612 (2016) 110

\bibitem{HKLL}
  A. Hamilton, D. N. Kabat, G. Lifschytz, D. A. Lowe, ``Local bulk operators in AdS/CFT: A Boundary view of horizons and locality," Phys.Rev. D73 (2006) 086003

\bibitem{InsideOut}
  A. Almheiri, T. Anous, A. Lewkowycz, ``Inside out: meet the operators inside the horizon. On bulk reconstruction behind causal horizons," JHEP 1801 (2018) 028
	
\bibitem{ErrorCorrection}
  A. Almheiri, X. Dong, D. Harlow, ``Bulk Locality and Quantum Error Correction in AdS/CFT," JHEP 1504 (2015) 163

\bibitem{HyperbolicScrambling}
  Y. Ahn, V. Jahnke, H.-S. Jeong. K.-Y. Kim, ``Scrambling in Hyperbolic Black Holes: shock waves and pole-skipping," arXiv:1907.08030 [hep-th]

\bibitem{BTZDynamics}
  R. R. Poojary, ``BTZ dynamics and chaos," arXiv:1812.10073 [hep-th]
  
\bibitem{ChaosBoundRotating}
  V. Jahnke, K.-Y. Kim, J. Yoon, ``On the Chaos Bound in Rotating Black Holes," JHEP 1905 (2019) 037

\bibitem{ComplexityGeometry}
  H. W. Lin, L. Susskind, ``Complexity Geometry and Schwarzian Dynamics," arxiv:1911.02603 [hep-th]

\bibitem{BulkSYK}
  Y. Lensky, X.-L. Qi, ``Operator size distribution and the holographic dual of SYK model," \textit{in preparation}



\end{thebibliography}

\end{document}